\documentclass[%
 aip,
 amsmath,amssymb,
 reprint,%
]{revtex4-1}
\usepackage{xcolor}
\usepackage{graphicx}
\usepackage{dcolumn}
\usepackage{bm}
\usepackage{setspace}
\usepackage[utf8]{inputenc}
\usepackage[T1]{fontenc}
\usepackage{mathptmx}
\usepackage{etoolbox}
\usepackage{amsmath} 
\usepackage{hyperref}
\usepackage{array}
\usepackage{amsmath, amssymb, xcolor}
\makeatletter
\def\@email#1#2{%
 \endgroup
 \patchcmd{\titleblock@produce}
  {\frontmatter@RRAPformat}
  {\frontmatter@RRAPformat{\produce@RRAP{*#1\href{mailto:#2}{#2}}}\frontmatter@RRAPformat}
  {}{}
}%
\makeatother

\newcommand{\btau}{\mathbf{\color{red}\tau}}

\begin{document}

\preprint{AIP/123-QED}

\title[ ]{A unifying perspective on measuring transient planar extensional viscosity from exponential shear}
\author{L.A. Kroo}
\affiliation{Massachusetts Institute of Technology, Department of Mechanical Engineering}
 \altaffiliation[L.K. Current affiliation: ]{Polymer Science and Engineering Department, University of Massachusetts Amherst}
\author{R.A. Nicholson}%
\author{M.W Boehm}%
\author{S.K. Baier}%
 \email{lkroo@umass.edu}
 \email{gareth@mit.edu}
\affiliation{ 
Motif FoodWorks
}%

\author{G.H. McKinley}
\affiliation{Massachusetts Institute of Technology, Department of Mechanical Engineering}

\date{\today}
\begin{abstract}
Here we present an experimentally practical and robust method to compute the transient extensional viscosity from exponential shear on a wide variety of viscoelastic complex fluids. To achieve this, we derive an analytical, frame-invariant continuum solution for the exponential shear material function ($\eta^+_{ES}$), valid over all Weissenberg numbers. Specifically, we amend the original framework of Doshi and Dealy proposed in 1987 to explicitly address the effect of rotation of material elements, due to the presence of vorticity. Modern strain-controlled rheometers can access a wide range of effective Hencky strain rates in exponential shear (approximately within a range of 0.01 to $7$ s$^{-1}$) and up to an effective Hencky strain of approximately 6 -- sufficient range to observe finite extensibility for many polymeric fluids. We quantify the kinematic, transducer-based, and instability-related experimental limitations of the method, establishing firm windows of data validity.  

The new material function is tested on a number of different example fluids. We show that these exponentially increasing strain histories are capable of producing "strong flow" -- generating stress-growth dynamics consistent with coil-stretch conformational changes in polymer solutions exhibited in extensional flows ($\mathrm{Wi}  >0.5$). We then demonstrate that the method can reach finite extensibility for a 0.3\% wt. PIB solution. The method is then quantitatively validated (with no fitted parameters) against a traditional extensional technique, Capillary Break-Up Extensional Rheometry (CaBER) using the same PIB solution. Questions related to generality of this approach are addressed, including discussions on multi-mode relaxation and shear thinning.

\end{abstract}

\maketitle

\begin{quotation}

This article guides the reader on how to use a torsional rheometer to quantitatively measure the transient planar extensional viscosity of a complex fluid. To accomplish this, we propose a Weissenberg-number dependent update to the seminal method of "Exponential Shear", that requires only experimental data. 




\end{quotation}

\hspace{-30pt}
\begin{table*}[!ht]
\caption{Nomenclature}
    \centering
    \small

    \begin{tabular}{cllll}
    \toprule
         \textbf{} & \textbf{Nomenclature} & \textbf{Form} & \textbf{Units} & \textbf{Time-Dependent?} \\
         \hline
                 
\textbf{} & \textbf{} & \textbf{} & \textbf{} & \textbf{} \\

$\textbf{Wi}$ & Weissenberg number & $\dot{\varepsilon}\tau$ & [-] ~ ~~~~~~~ & Constant\\
$\textbf{B}$ & Finger strain tensor & -- & [-] ~ ~~~~~~~ & Time-varying (prescribed)\\
$I_1,I_2$ & Strain invariants & -- & [-] & Time-varying (prescribed)\\
$\gamma$ & Strain & -- & [-] & Time-varying (prescribed)\\
$\dot{\gamma}$ & Strain rate & -- & [s$^{-1}$] & Time-varying (prescribed)\\
$\chi $ & Kinematic angle of strain & -- & [rad] & Time-varying (prescribed)\\
$\theta $ & Angle of principal axes of stress tensor & -- & [rad] & Time-varying\\
$\lambda_a$ & Principal stretch of a material element & -- & [-] & Time-varying (prescribed) \\
$\varepsilon$ & Hencky strain (extension) & $\ln(\lambda_a(t))$ & [-] & Time-varying (prescribed)\\
$\dot{\varepsilon}$ & Hencky strain rate (extension) & $\dot{\lambda}_a(t)/\lambda_a(t)$ & [s$^{-1}$] & Constant (prescribed)\\
$\mu$ & Steady-state Newtonian viscosity & $\sigma_{yx} / \dot{\gamma}_{yx} $ & Pa$\cdot$s & Constant\\
$\eta_{p}$ & Polymeric contribution to the shear viscosity & -- & Pa$\cdot$s & Time-varying \\
$\eta_{s}$ & Solvent contribution to the shear viscosity & -- & Pa$\cdot$s & Constant \\
$\eta_{PE}^+$ & Transient planar extensional viscosity & $\Delta \sigma / \dot{\varepsilon} $& Pa$\cdot$s & Time-varying\\
$\eta_{ES}^+$ & Exponential shear material function &$\Delta \sigma / RF$ & Pa$\cdot$s & Time-varying \\
$SR$ & Stress ratio & $N_1/(2\sigma_{yx})$~~~~~ & [-] & Time-varying \\
$RF$ & Rate factor or "effective stretching rate" in exponential shear &$RF \rightarrow \alpha$ for Wi $\gg1$ ~~~~~ & [s$^{-1}$] & Time-varying ($0 < Wi\lesssim 3$) OR constant ($Wi\gtrsim 3$) \\
$\tau$ & Relaxation time of fluid (UCM or LVE)& -- & [s] & Constant \\
$t^*$ & Proposed "stress-growth" time constant & $\dfrac{N_1(t^*)}{2 \sigma_{yx}(t^*)}$ =1 & [s] & Constant \\
$\eta_0$ & Zero-shear total viscosity of fluid & -- & Pa$\cdot$s & Constant \\
$G$ & Elastic modulus of fluid & $\eta_0 / \tau$ & [Pa] & Constant \\

\textbf{} & \textbf{} & \textbf{} & \textbf{} & \textbf{} \\

\toprule
\end{tabular}
\label{tab:my_label}
\end{table*}


\section{\label{sec:level1}Introduction :\protect\\}
\subsection{\label{sec:history} Brief History: Exponential Shear}

The transient \textit{extensional} properties of complex fluids are notoriously difficult to measure experimentally, yet critical to industrial processing of polymeric complex fluids. In 1987, a novel and convenient method was proposed to measure the transient planar extensional viscosity ($\eta^+_p$) of complex fluids -- but using a standard strain-controlled \textit{torsional} rheometer (rather than a specialized extensional rheometer). This idea, first proposed by Doshi and Dealy was called \textit{"exponential shear"}\cite{doshi1987exponential}. 

Over the past 50 years, exponential shear has become a controversial method in the field of rheology - with evident validity in some cases \cite{kwan1999brownian, kwan2001experimental, graham2001pom,wagner2005exponential,suneel2003characterisation}, and apparent failure in others \cite{venerus2000exponential,neergaard2000exponential,samurkas1989strong}. Due to this conflict in the literature, academia and industry have both used caution and avoided this measurement technique since about 2005\cite{wagner2005exponential}, instead primarily focusing on the development of novel extensional rheometers for these highly non-linear dynamical measurements (examples include the Meissner apparatus\cite{meissner1972development}, SER \cite{sentmanat2005measuring}, Filament-stretching \cite{mckinley2002filament}, elasto-capillary methods \cite{anna2001elasto, dinic2017pinch}, and microfluidic methods such as opposed contraction jet devices \cite{pipe2007microfluidic}, Optimized Shape Cross-slot Extensional Rheometers (OSCER)\cite{haward2023extensional2}) and Optimized uniaxial and biaxial extensional rheometers (OUBER)\cite{haward2023extensional1}. 

 Despite significant progress in methods development and a fast-growing academic community now making scientific progress in understanding the non-linear dynamics of complex fluids in stretching flows \cite{doyle1998relaxation, schroeder2003observation}, most extensional rheometers remain highly specialized instruments, each with its own set of somewhat restrictive limitations. For example, some of these limitations include gravity effects, interfacial dynamics, solvent evaporation from the large interfaces, or in the case of uniaxial elasto-capillary-driven devices, the strict restriction to a specific constant Hencky strain rate\cite{entov1997effect}, such that Wi = 2/3. In this investigation, we return to this historical puzzle of exponential shear, in an effort to elucidate the conflicting findings of prior studies, and make accessible a method to accurately employ exponential shear broadly on complex fluids across a wide range of stretching rates relevant to processing industries (for example in polymer processing, food science, etc.). 
 
\begin{figure}[ht!]
\includegraphics[width=\linewidth]{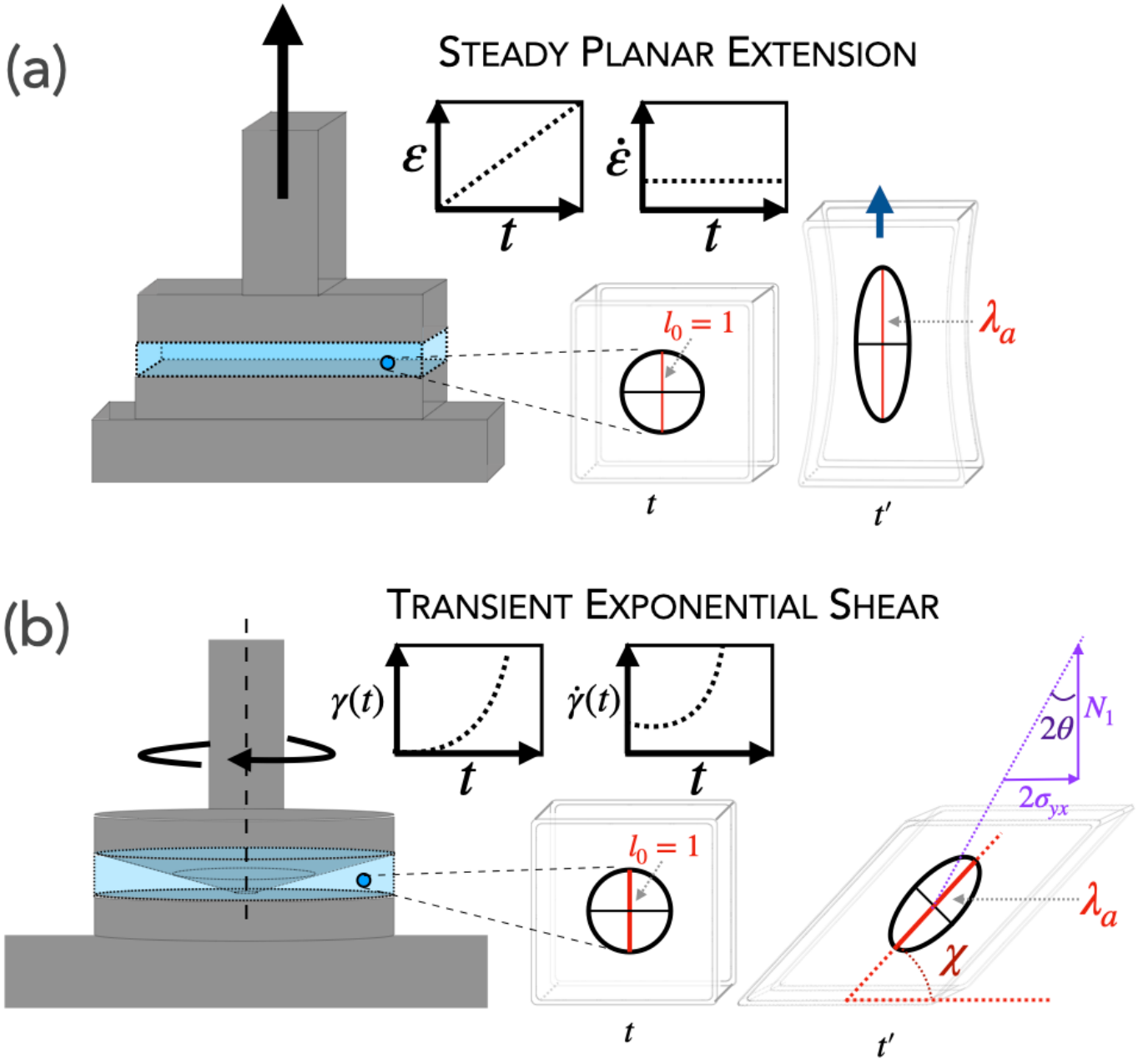}
\caption{\label{fig:concept} Figure 1: (a) Steady Planar Extension is depicted on a fluid sample, where the stretch of a bulk material element,$\lambda_{a}$ , is a kinematic quantity set by the imposed Hencky strain at a constant Hencky strain rate.  (b) When a specific transient shear strain is imposed in a cone-and-plate geometry,  (“exponential shear”), the stretch of a material element, $\lambda_{a}$is identical to the stretch of an element in constant-rate steady planar extension (shown in a). However, in the case of exponential shear, the element is also rotated by the kinematic angle $\chi$.This angle is set by the magnitude of the strain amplitude, $\gamma$. The angle of the maximum stresses (found via measured quantities) is separately defined as $\theta$ in purple, as shown.  }
\end{figure}

\subsection{\label{sec:ES} The Technical Concept: Exponential Shear}

The idea originally presented by Doshi and Dealy in 1987\cite{doshi1987exponential} was that in a steady planar extensional flow (see Figure \ref{fig:concept}a), material elements are stretched ($\lambda_{PE}$) along the axis of principal stress. Material elements in shear flow also elongate along an axis, but at some evolving angle, leading to a similar effect where material elements are also stretched ($\lambda_{ES}$). The idea of exponential shear is to impose a specific \textit{transient} strain function such that the stretch of material-elements (in shear versus extension) are exactly equal at all times: $\lambda_{ES}(t) = \lambda_{PE}(t)$. 

Polymeric fluids in stretching flows undergo a conformational change, dynamically unfolding from an equilibrium "coiled" state to a "fully stretched", rod-like state when in a sufficiently strong flow\cite{de1974coil}. This dynamic, transient morphing of microstructure typically causes strain-rate hardening in the bulk, where the bulk extensional viscosity undergoes dramatic, nonlinear stiffening (often by many orders of magnitude)\cite{munstedt1979elongational}. After this period of stress growth, the transient extensional viscosity saturates in real polymeric fluids\cite{kubinski2024extensional} when they are fully extended -- at a "finite extensibility" of the molecule or microstructure. By treating transient extensional viscosity as a \textit{proxy} for ensemble-averaged molecular conformation, we can rapidly infer information about some of the hardest-to-measure, dynamical behavior of molecules -- without expensive imaging tools or specialized extensional rheometers. 

Exponential shear could enable laboratories to perform such measurements by using common, bench-top equipment (i.e. a strain-controlled torsional rheometer), enabling key measurements that are crucial (and currently inaccessible) to numerous industries. 


\section{\label{sec:level1} Kinematics \protect\\ }
In brief review, "exponential shear" is defined as the strain function that identically matches the stretch and stretch rates of a steady extensional flow ($\dot{\varepsilon} = \alpha =$ constant). We revisit this derivation below: 

In planar extension (i.e. $v_x = \dot{\varepsilon}x$,~ $v_y = -\dot{\varepsilon}y$, ~ $v_z = 0 $, as shown in Fig 1a), the stretch of a material element can be written: 
\begin{equation}
    \lambda_{a,(ext)} = e^{\dot{\varepsilon t}}
\end{equation}

In shear, the stretch of a material is related to the kinematic angle of the material element, $\chi(t)$: 
 \begin{equation}
    \lambda_{a,(shear)}  = \cot(\chi) 
\end{equation}
The angle, $\chi$, is prescribed instantaneously / geometrically by the imposed strain amplitude, in the "affine" limit (as shown in red in Fig 1b): 
 \begin{equation}
    \gamma(t)  = 2 \cot{2 \chi} 
\end{equation}
This naturally gives rise to the stretch in shear (given first by Lodge \cite{lodge1964elastic}):
\begin{equation}
\label{eq:lodgeStretch}
    \lambda_{a,(shear)} = \dfrac{\gamma(t) + \sqrt{\gamma(t)^2+4}}{2} 
\end{equation}

If eq \ref{eq:lodgeStretch} is set equal to eq 1, it is evident that the proper strain function, $\gamma(t)$, that will equate the stretches is precisely: 

\begin{equation}
\label{eq:ESstrain}
    \gamma(t) = e^{{\alpha}t } - e^{-{\alpha}t} = 2 \sinh(\alpha t),
\end{equation} where $\alpha = \dot{\varepsilon}$, the effective Hencky strain rate in exponential shear. 

Critical to this derivation: the "stretch" of a material element is interpreted in this derivation as a \textbf{purely kinematic quantity}; something entirely prescribed by the boundary conditions imposed by a strain-controlled deformation. This is important to delineate from micro-structural stretch in complex fluids (or chain-stretch in polymer solutions). 


\subsection{\label{sec:flowtype}"Strong Flow" and Flow type}

\begin{figure}[ht!]
\includegraphics[width=0.8\linewidth]{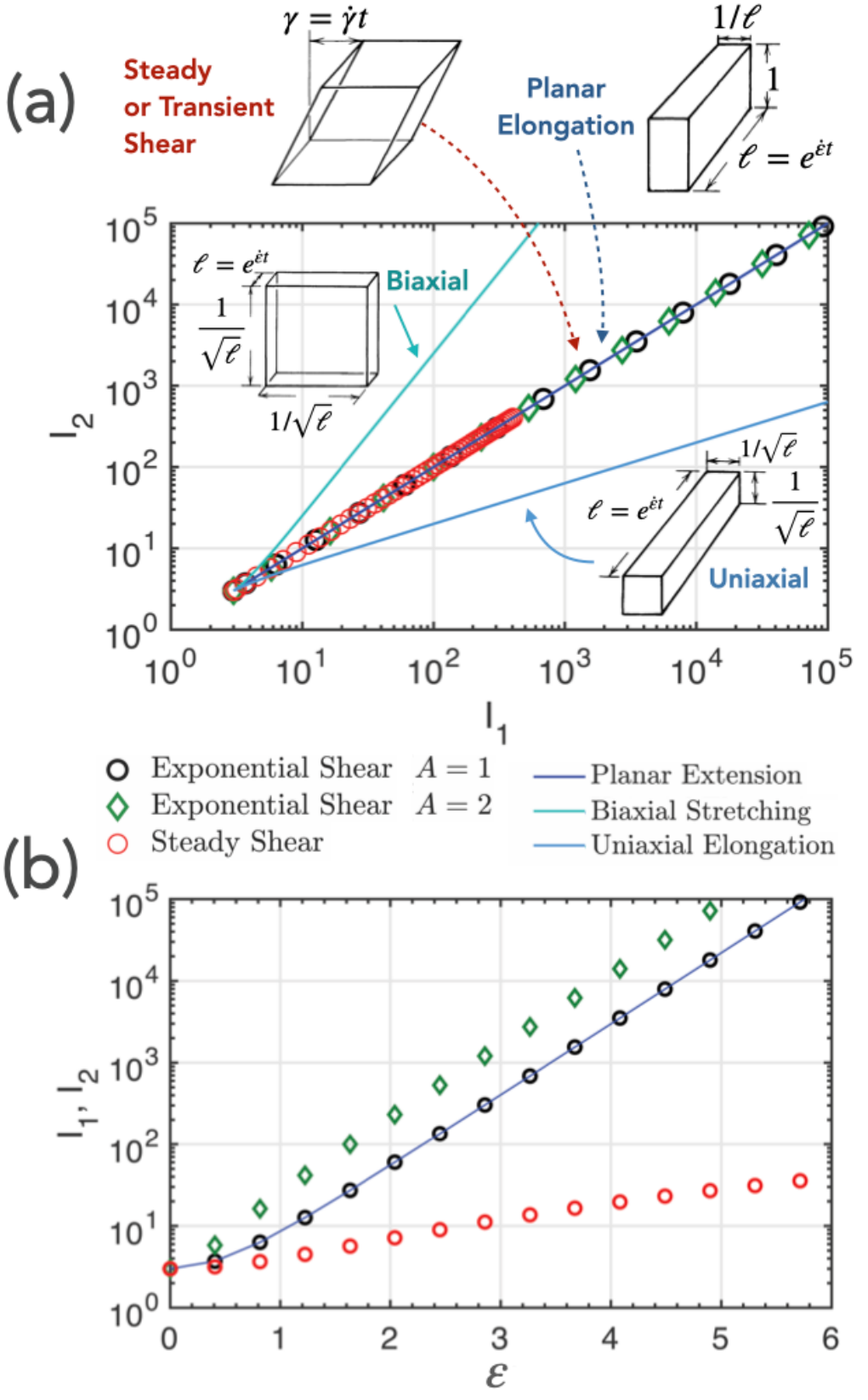}
\caption{\label{fig:invariants}(a) The kinematic equivalence between exponential shear and planar extension can also be shown in the strain invariants space. An incompressible fluid in a steady or transient shear history traverses the same set of “states” as planar elongation. Here we plot the first and second strain invariants for different flow types and for steady versus transient shear— inspired by the original framework by \citet{rivlin1997strain}. (b)  However, only one specific transient strain history accurately mimics steady planar extension in time (or strain). Here we show visually what is derived in the text: the transient function  matches steady planar extension identically in time.  }
\end{figure}
This resulting strain function (eq 6) was first posed by Doshi and Dealy in their seminal paper in 1987 \cite{doshi1987exponential}, with an additional constant, "$A$", as an unknown leading coefficient: $$\gamma(t) = A (e^{{\alpha}t } - e^{-{\alpha}t})$$. 
The rationale behind the choice of constant "A" was not previously addressed in those seminal papers, although typically set to unity in following work. This tunable constant perhaps reflected some uncertainty related to the relevant extensional flow type - which was a later topic of debate in the literature. Kwan and Shaqfeh \cite{kwan2001experimental} showed remarkable similarity with either planar or uniaxial extension at high Wi. At high strain amplitude, extensional techniques for either converged anyway.  
However, we argue here that the appropriate choice of the constant is $A = 1$  -- and that the equivalent flow-type is indeed planar extension (not uniaxial). From a frame-invariant perspective, we can compute the invariants of the Finger strain tensor, $\textbf{B}$ (Recall: $\textbf{B} = F \cdot F^t$). $I_1$, $I_2$, and $I_3$ represent a low-dimensional of the total stretch of a material, the total shape deformation of the material, and the compressibility of the material, irrespective of rotations, flow type or coordinate frame motion. As depicted in Fig \ref{fig:invariants}b, planar extension and any transient shear flow will transverse the same line in this state space of $I_1$ and $I_2$.  
If we consider the trajectories of $I_1$ and $I_2$ in time (or as a function of accumulated Hencky strain, $ \varepsilon = \alpha t =\dot{\varepsilon t}$, it is evident that $A = 1$ is a \textit{necessary} condition (equating \textcite{doshi1987exponential}'s original strain function to eq \ref{eq:ESstrain}). This strain function is shown in Figure \ref{fig:invariants}b, up to a Hencky strain of $\varepsilon = 6$ (in black circle symbols) is equal to planar extension (solid line).   

\subsection{\label{sec:PS} Defining Total Principal Stress in Extension and Shear}

One key challenge in exponential shear is to use experimental data (without fits or model parameters) to define a material function for exponential shear that exactly matches the planar extensional viscosity, 

\begin{equation}
\label{define}
    \eta_{ES}^+(\alpha t) = \eta_{PE}^+(\dot{\varepsilon t})
\end{equation}

A number of conflicting solutions for $\eta_{ES}^+(\alpha t)$ have been proposed in past literature. To approach the question of validity of these solutions (in section IIc), we start by considering the Upper Convected Maxwell constitutive equation, as an \textit{example} non-linear viscoelastic fluid: 
\begin{equation}
\label{eqn:UCM}
\boldsymbol{\sigma} + \tau \overset{\nabla}{\boldsymbol{\sigma}} = \eta_p \dot{\boldsymbol{\gamma}}
\end{equation}

where the upper convected derivative is given by: $
\overset{\nabla}{\boldsymbol{\sigma}} = \dfrac{D\boldsymbol{\sigma}}{Dt} - \left[ (\nabla \mathbf{v})^T \cdot \boldsymbol{\sigma} + \boldsymbol{\sigma} \cdot (\nabla \mathbf{v}) \right]
$. This single-mode, nonlinear model fluid provides synthetic data to rigorously test the previously proposed material functions against for validity in different conditions.

For a planar extensional flow, extending along $x$, the transient extensional viscosity is defined as: 
\begin{equation}
    \eta_p^+ (\dot{\varepsilon} t) = \frac{\sigma_{xx} - \sigma_{yy}}{ \dot{\varepsilon}} = \dfrac{\Delta \sigma_{PE} }{ { \dot{\varepsilon}}}
\end{equation}

It is straightforward to solve eq 7 for $\sigma_{xx}$ and $\sigma_{yy}$, assuming $\mathbf{\dot{\gamma}} = 
\begin{bmatrix}
\dot{\varepsilon} & 0 & 0 \\
0 & -\dot{\varepsilon} & 0 \\
0 & 0 & 0
\end{bmatrix}$, and compute the total principal stress for a UCM fluid in planar extension: 

\begin{equation}
\label{eq:planar}
    \Delta \sigma_{PE}(\varepsilon) \;=\;
\eta_p \dot{\varepsilon} \Biggl[
  \dfrac{\,1 - e^{\!\left(-\frac{(1 - 2\,\tau\,\dot{\varepsilon})}{\tau}\,t\right)\,}}
       {\,1 - 2\,\tau\,\dot{\varepsilon}\,}
  \;+\;
  \frac{\,1 - e^{\!\left(-\frac{(1 + 2\,\tau\,\dot{\varepsilon})}{\tau}\, t \right)\,}}
       {\,1 + 2\,\tau\,\dot{\varepsilon}\,}
\Biggr] \\ 
\end{equation}
\vspace{5pt}
For different values of Weissenberg number (Wi), where $ \mathrm{Wi} \equiv \dot{\varepsilon} \tau$, we can visualize the response of the total principal stress in an ideal extension, as shown in figure \ref{fig:UCMresult}b. 

Notice at small Wi ($Wi \ll 0.5$), the principal stress reaches a steady state. In the moderate Wi regime ($Wi \approx O(1)$) , the fluid no longer reaches a steady state - locked in a domain where the stress-growth is faster than the fading memory of the fluid. This is arguably the continuum-mechanics definition of a "strong flow": a flow that causes  stress growth to outpace fading memory. Note that even for very high Wi ($> O(10)$), this continuum model does not capture finite extensibility (a later plateau). From a molecular perspective, many studies have previously shown that this transition from low Wi to high Wi also corresponds with the ability for the flow to coil-stretch molecules.

\begin{figure*}[ht!]
\label{fig:UCMresult}
\includegraphics[width=6.1 in]{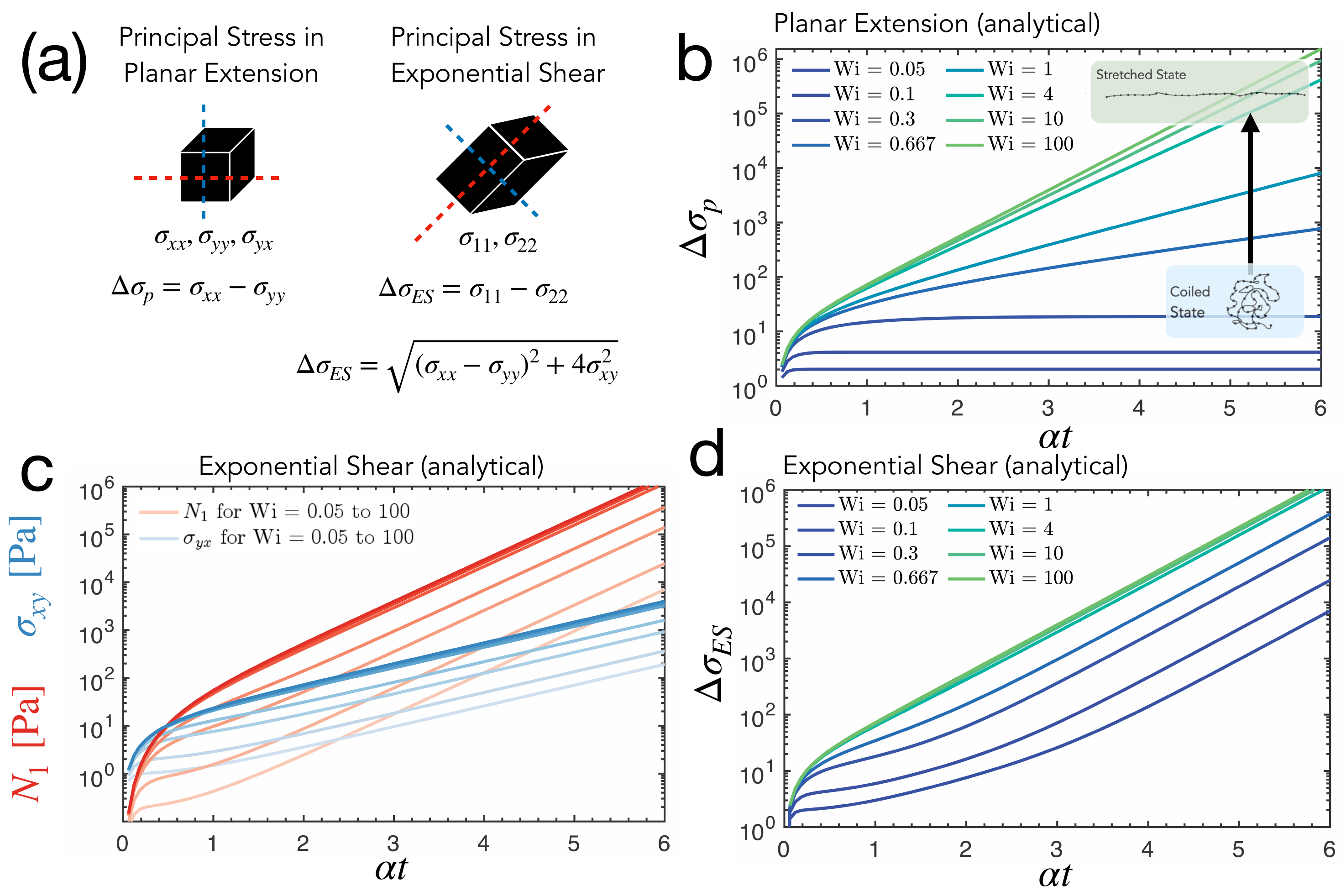}[ht!]
\caption{\label{fig:UCMresult}Figure 3: (a) The total principal stress (defined as the magnitude of the maximum stress) in the material element is shown for planar extension (left) versus exponential shear (right). As shown in the text,  is derived by rotating the stress tensor to the angle which maximizes the magnitude of the stress (by an angle, $\theta$, shown in Fig 1). Interestingly, these definitions do not assume that angle of maximum stress is collinear with the kinematic angle of stretch (i.e. $\chi $ and $ \theta$ may not be equal in all cases). (b) The total principal stress is shown for planar extension for an idealized model fluid (Upper-Convected Maxwell fluid) for a range of different Wi. (c) Synthetic data is shown for the first normal stress difference and shear stress (two independently measured components) for a range of Wi for a UCM fluid in exponential shear (d) The total principal stress is shown for exponential shear using the same constitutive model. }
\end{figure*}

Because this constitutive model is frame-invariant and can be solved in any flow-type -- we can apply the same procedure to solve for the stresses (particularly, the \textit{total principal stress}) in an exponential shear flow. 

Eq 7 can be rewritten as a coupled set of ordinary differential equations: 

\begin{eqnarray}
    \frac{d \sigma_{yx}}{d t} = \frac{\eta_p}{\tau} \dot{\gamma} - \frac{\sigma_{yx}}{\tau} \\
    \frac{d \sigma_{xx}}{d t} = 2 \sigma_{yx} \dot{\gamma} - \frac{\sigma_{xx}}{\tau} \\
    \sigma_{yy} = 0
\end{eqnarray}

Assuming the exponential shear strain function is given by $\dot{\gamma}_{yx}(t) = 2 \alpha \cosh(\alpha t)$,  and a homogeneous strain field, 
$\dot{\boldsymbol{\gamma}}(t) =
\begin{bmatrix}
0 & \dot{\gamma}_{yx}(t) & 0 \\
\dot{\gamma}_{yx}(t) & 0 & 0 \\
0 & 0 & 0
\end{bmatrix})$, the solution for the time-evolving shear stress is:

\begin{equation}
\label{eqn:sigmayxUCM}
    \sigma_{yx} = \frac{\eta_p \alpha}{\alpha^2 \tau^2 -1} \left( \alpha \tau \gamma - \frac{\dot{\gamma}}{\alpha} + 2 e^{-t/\tau}\right).
\end{equation}
<NOT: add weissenberg discussion>
<more discussion on the 3 terms>

This solution is shown in blue in Figure \ref{fig:UCMresult}c for a large range of Wi. 

The solution for the (more aggressively evolving) normal stress difference ($\sigma_{xx}-\sigma_{yy}$) is: 
\begin{equation*}
    N_1 = \frac{2\alpha^2 \tau \eta_p}{\alpha^2 \tau^2 - 1 } \mathrm{\bigg (}  \frac{2 \alpha^2\tau^2+1}{4 \alpha^2 \tau^2 -1}  \left( \frac{\gamma^2}{2} + \frac{\dot{\gamma}^2}{2 \alpha^2} \right) ... \end{equation*}
\begin{equation}
\label{eq:N1}
    -\frac{3 \tau \gamma \dot{\gamma}}{4 \alpha^2\tau^2-1} +
\frac{2 \gamma e^{-t/\tau}}{\alpha \tau} -
\frac{3 e^{-t/\tau}}{4 \alpha^2 \tau^2 -1} +
e^{-t/\tau} -2 \mathrm{\bigg ).}
\end{equation}

This normal stress difference is also shown on Figure \ref{fig:UCMresult}c (in red), and grows faster than the shear stress. 

We note that after a systematic asymptotic analysis of Eq. 14 that a (non-trivial) simplification is robust over all Wi (but at intermediate to large Hencky strains) can be given by:

\begin{equation}
    {N_1} = \frac{\eta_p}{\tau} \left[ \frac{2 \, (\alpha \tau)^2}{(\alpha\tau + 1)(2 \, \alpha\tau + 1)} \right] e^{2 \alpha t}
\end{equation}

The total principal stress in shear, $\Delta \sigma_{ES}$, is conceptually the stress along the angle that maximizes the stress (a combination of both $N_1$ and $\sigma_{yx}$. See fig 3a). The standard deviatoric stress tensor in a (typical) complex fluid can be written:

\begin{equation}
\label{eq:stressTensor}
\boldsymbol{\sigma} =
\begin{pmatrix}
\sigma_{xx} & \sigma_{xy} & 0 \\
\sigma_{xy} & \sigma_{yy} & 0 \\
0          & 0          & \sigma_{zz}
\end{pmatrix}
\end{equation}

The eigenvalues, $\Lambda$ of eq. \ref{eq:stressTensor} must satisfy: 
\begin{equation}
\det\bigl(\boldsymbol{\sigma}-\Lambda\mathbf{I}\bigr)=0
\end{equation}

Evaluating the determinant and solving for the two non-zero eigenvalues:
\begin{equation}
\Lambda_{1,2}=\frac{\sigma_{xx}+\sigma_{yy}}{2}\pm\sqrt{\left(\frac{\sigma_{xx}-\sigma_{yy}}{2}\right)^2+\sigma_{xy}^2}
\end{equation}

Simplifying in terms of $N_1$ and $\sigma_{yx}$, the total principal stress is given by:
\begin{equation}
\label{eq:DS}
\Delta \sigma=\Lambda_1-\Lambda_2=2\sqrt{\left(\frac{N_1}{2}\right)^2+\sigma_{xy}^2} = \sqrt{N_1^2+4\sigma_{yx}^2}
\end{equation}

This is equivalent to manually solving for the rotation of the stress tensor to an angle, $\theta$, that maximizes the stress (i.e. $
\sigma_n(\theta)=\sigma_{xx}\cos^2\theta +2\sigma_{xy}\sin\theta\cos\theta+\sigma_{yy}\sin^2\theta$ and solving $\frac{d\sigma_n}{d\theta} = 0$). 

The angle that aligns the stress tensor with the principal axes is also useful, and can be written:
\begin{equation}
\theta = \frac{1}{2} \cot^{-1}(\frac{N_1}{2 \sigma_{yx}}),
\end{equation}
as observed in prior work \cite{doshi1987exponential,kwan1999brownian}. This angle conceptually is depicted in purple in Fig 1b, and \textit{can be} a distinct angle from $\chi$, the kinematic angle of stretch. 

Substituting eq. \ref{eqn:sigmayxUCM} and \ref{eq:N1} into eq. \ref{eq:DS}, we can see that the total principal stress of an Upper-Convected Maxwell fluid takes the somewhat complicated form:
\newpage
\begin{eqnarray*} 
\Delta\sigma_{ES} = \frac{2\alpha}{\alpha^2\tau^2-1}
\mathrm{\bigg[ }
  \alpha^2\tau^2\eta_0^2
   \mathrm{\bigg[}
    \frac{2\alpha^2\tau^2+1}{4\alpha^2\tau^2-1}\left(
      \frac{\gamma^2}{2} + \frac{\dot{\gamma}^2}{2\alpha^2}
    \right)  \\
    - \frac{3\tau\gamma\dot{\gamma}}{4\alpha^2\tau^2-1}
     + \frac{2\gamma e^{-t/\tau}}{\alpha\tau}
    - \frac{3e^{-t/\tau}}{4\alpha^2\tau^2-1}
    + e^{-t/\tau} - 2 \mathrm{\bigg]}^2
  \end{eqnarray*}
  \vspace{-15pt}
  \begin{equation}
  + \eta_p^2\left(
    \alpha\tau\gamma - \frac{\dot{\gamma}}{\alpha} + 2e^{-t/\tau}
  \right)^2
\mathrm{\bigg ]}^{1/2}
\end{equation}

This function is plotted in Figure \ref{fig:UCMresult}d. As shown, the principal stress in exponential shear appears identical to the evolution of the principal stress in planar extension at high Wi. However, $\Delta \sigma_{ES}$ systematically deviates from $\Delta \sigma_{PE}$ at low and intermediate Wi.  

\subsection{\label{sec:materialfunction} The Material Function Debate: none "work" at moderate Wi}
\begin{figure}
\includegraphics[width=0.9 \linewidth]{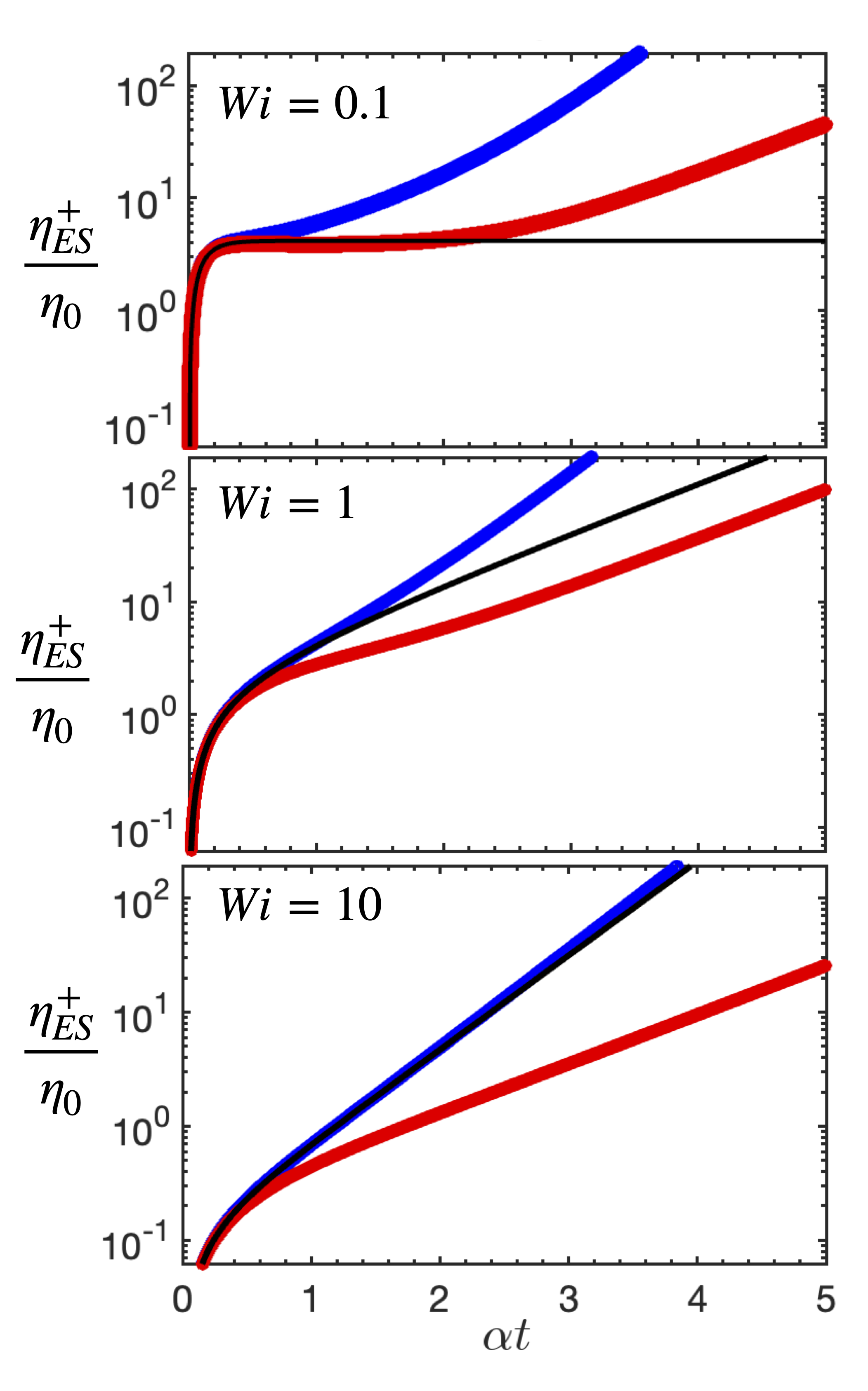}
\caption{\label{fig:4} Figure 4: Here we compare the analytical solution for planar extension (for an Upper-Convected Maxwell fluid) with the two previous material functions previously proposed by the community. Blue lines correspond to the material function favored by Kwan and Shaqfeh. Red lines correspond o the material function proposed originally by Doshi and Dealy. Black lines are true planar extension. The ideal material function would be able to exactly replicate the planar extension over all Wi, and over all strain amplitudes. Note that neither of the two popular material functions can quantitatively capture the transient extensional viscosity at intermediate Wi.   }
\end{figure}
An extensional viscosity material function is defined as a principal stress (i.e. the stress along a principal axes of material stretch) divided by a Hencky strain rate (in extension, this rate is $ \dot{\varepsilon} = \frac{d{\ln\lambda}}{dt}$. 

In exponential shear, the \textit{effective} Hencky strain rate is a subject of contentious debate. Several candidates for a material function have been proposed and studied that offer different perspectives on the correct rate. The two most accepted material functions include one by Doshi and Dealy :
\begin{equation}
\label{matFuncDoshi}
    \eta_{ES}^+ = \frac{\sqrt{N_1^2 + 4 \sigma_{yx}^2}}{2 \dot{\gamma}(t)}
\end{equation}

An alternative was investigated by Kwan and Shaqfeh: 
\begin{equation}
\label{eq:Kwan}
    \eta_{ES}^+ = \frac{\sqrt{N_1^2 + 4 \sigma_{yx}^2}}{\alpha}
\end{equation}

 Various other forms (with simplified numerators) have also commonly used by Mcleish, Larson, and others including $\eta_{ES}^+ = \dfrac{N_1}{\alpha}
$ and $\eta_{ES}^+ = \dfrac{\sigma_{yx}}{\dot{\gamma}}$.

Using synthetic data from our Upper-Convected Maxwell model, we tested these material functions (Eq \ref{matFuncDoshi}, and Eq \ref{eq:Kwan}) -- to compare against the simulated planar extensional viscosity (Eq \ref{eq:planar}) at different Wi. As shown in figure \ref{fig:4}, Eq \ref{eq:Kwan} accurately approaches $\eta_{PE}^+$ at high Wi. Eq \ref{matFuncDoshi} converges to $\eta_{PE}^+$ at very low Wi. At intermediate Wi -- arguably the most critical to characterize from an experimental perspective -- \textit{neither} function can quantitatively capture the transient extensional viscosity at finite Hencky strain. 

Because up to this point, the UCM model is analytical and self-consistent -- this points to a fundamental issue with exponential shear that is deeper than experimental limitations.

Based on figure \ref{fig:4}, the only material function that is reasonably robust is Eq \ref{eq:Kwan} ($\eta_{ES}^+ = \frac{\sqrt{N_1^2 + 4 \sigma_{yx}^2}}{\alpha}$ ) in the strict limit that  Wi$ \gg 1$. However, to know the Wi$ \gg 1$, the experimentalist must know (roughly) the relaxation time of the fluid. If studying aggressive extensional flows at large Hencky strain rates (such as for extreme processing conditions), this material function may be sufficient. Such experiments, however, cannot be validated against elasto-capillary extensional rheometry methods (as these methods can only access $Wi = 2/3$ in uniaxial extension).

\subsection{Recoverable strain and non-affinity in Complex Fluids at intermediate to low Wi}
\label{sec:recoverableStrain}
\begin{figure*}[ht!]
\includegraphics[width=0.7\linewidth]{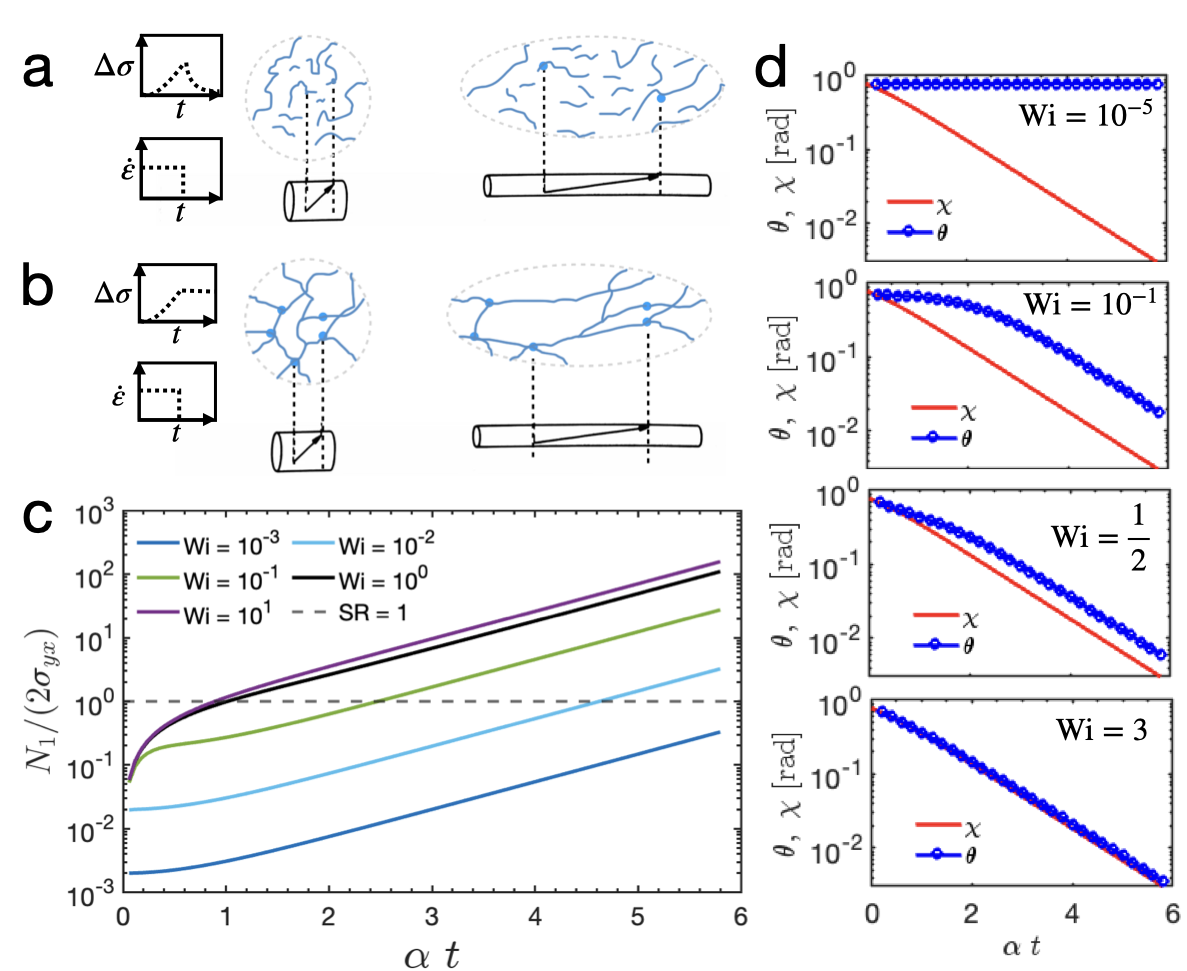}
\caption{\label{fig:stressRatio} (a) Example of a "non-affine" system under planar extension. In this linear viscoelastic case, the total principal stress at the cessation of extension would decay as $\lim_{\alpha\tau \to \ll1} \Delta \sigma = 4 G \mathrm{Wi} (1-e^{-t/\tau})$. (b) Example of an "affine" hyper-elastic nonlinear solid in extension, where the stress would not decay at all, but hold where $\lim_{\alpha\tau \to \infty} \Delta \sigma = G(\lambda_a^2-\lambda_a^{-2})$ (c) The evolving recoverable strain or "stress ratio", as described in eq \ref{eq:stressRatio} at different Wi for a UCM fluid (eq \ref{eqn:UCM}). (d) The angle of the principal of the stress ($\theta$, in blue) versus the strain ($\chi$, in red) are depicted for increasing Wi. At small Wi (Newtonian limit), these angles are completed uncoupled. At the high Wi (affine limit), these angles are equivalent. 
}
\end{figure*}

The key conceptual \textit{reason} that neither of the proposed material functions in figure \ref{fig:4} represent planar extension for intermediate, finite Weissenberg number is related to the original assumption that was used to derive equation \ref{eq:lodgeStretch}: that stretch is a controlled quantity by the imposed transient strain, irrespective of the fluid microstructure or internal stresses. 

In figure \ref{fig:stressRatio}d, we can see that the principal axes of $\chi$ and $\theta$ are not necessarily collinear -- indeed they only are colinear in the limit of Wi$\gg 1$. Kwan and Shaqfeh\cite{kwan1999brownian} recognized this precise issue that $\chi$ and $\theta$ deviated at low and intermediate Wi, and originally suggested modification of the \textit{numerator} of the exponential shear material function with additional explicit, micro-structural angle information (which they suggested the user obtain through birefringence imaging) to correct for this. However, this function appeared to not address the issue suitably (as they attempted to correct the principal stress rather than the effective rate). Indeed, they concluded the proposed material function appeared equal to $\alpha$ even at intermediate and low Wi.   

The key assumption Lodge uses to derive eq \ref{eq:lodgeStretch} is that the microstructural elements (such as tube segments in reptation theory) must deform exactly as prescribed by the macroscopic deformation gradient, $\textbf{F}$, for eq 4 to reflect micro-structural stretch. In polymer theories, this is historically termed the "affine" limit\cite{larson2013constitutive}, valid for model materials with no convective constraint release (CCR) mechanism \cite{graham2003microscopic}. These "affine" material vectors would follow: $\delta \mathbf{r} = \mathbf{F} \cdot \delta \mathbf{r}_0$, and orientations would obey $  \mathbf{u}(t) = \dfrac{\mathbf{F}(t) \cdot \mathbf{u}_0}{\|\mathbf{F}(t) \cdot \mathbf{u}_0\|}$, without feedback from stresses. Given that the term 'affine' has recently been used more broadly in soft matter to describe phenomena related to strain localization (see, e.g., D2 metric), we will instead refer to this assumption that the principal axess of the stress and strain tensors are co-linear — \textit{as required by a purely kinematic description}— as the 'Controlled Stretch' assumption. 

Prior studies on flow classification and persistence of straining \cite{astarita1979objective,thompson2005persistence} point out that in real fluids with vorticity, stress is dissipated by the rotation of material elements; by microstructures tumbling.  Because the vorticity scales as: $\omega = \dfrac{1}{2}\dot{\gamma}(t)$, and shear stress also scales as $\dot{\gamma}$, there is a clear limit where exponential shear cannot generate enough stress to be considered a "strong flow".  However, these studies tend to examine the growth of the shear stress (rather than the total principal stress), neglecting the far more aggressive contribution of $N_1$ to the total stress, which grows faster than the vorticity term in strong flows (Wi $>0.5$). 

The simplest example of this case of a "non-affine" fluid is a simple Newtonian fluid ($\tau = 0$) in steady shear. In a Newtonian fluid, the principal stress direction remains fixed at 45$^\circ$ relative to the flow direction, regardless of the accumulated strain — a clear instance where the stress and strain tensors do not co-rotate, as would be expected under the affine, 'Controlled Stretch' assumption. 

In the specific case of cessation of a steady extension as depicted in (\ref{fig:stressRatio}a,b), the principal stresses will relax to different extents based on the degree of cross-linking. This is also a  well-studied idea for transient shear strain histories: perhaps the most explored example of this idea is cessation of steady shear on a stress-controlled rheometer. At the moment that the steady shear arrests - the stress (and strain) history are transient. Although no further stress is being applied, the fluid recoils or more accurately “recovers” in elastic materials due to entanglements, cross-linking, or more generally, \textit{any} micro-structural storage mechanism for elastic energy. 

In stress-controlled steady shear rheometry, this non-zero measured value of strain is called the “recoverable strain”\cite{singh2021revisiting}. This is a very simple case of where $\chi$ and $\theta$ are not collinear in a transient flow. 

This phenomena (of non-colinearity between stress and strain tensors) is naturally Weissenberg-number dependent. An essentially Newtonian fluid ($Wi\ll1$) will have almost no memory and will instantly forget its history — it will recover entirely and instantly upon cessation of shear. A soft solid in the Neo-Hookean limit will have an infinite memory, “freezing” its  microstructural configuration immediately upon cessation and maintaining internal stresses “affinely” with the imposed strain amplitude. In summary: an affine material is "stuck" at high Wi (no matter the imposed rate). An ideal Newtonian fluid is "stuck" at Wi = 0,  also irrespective of the imposed shear rate. A material with \textit{limited} recoverable strain would have some finite Wi, dependent on the rate of strain. In short: recoverable strain is related to the micro-structural capacity to \textit{store} elastic energy. That structural property that encodes memory also increases the coupling between shear stresses and normal stresses in a material. 

Indeed, Lodge\cite{lodge1958network} originally showed that the stress ratio:

  \begin{equation}
  \label{eq:stressRatio}
     SR =  N_1 / 2\sigma_{yx}
  \end{equation}

in shear flows is \textit{equal} to the magnitude of the recoverable strain. This has been used by \textcite{lee2019structure} on other large-amplitude shearing flows (LAOS, flow curves, etc.) to predict structure-property relationships in viscoelastic fluids. We have plotted this quantity for a UCM fluid in exponential shear in figure \ref{fig:4}c for different Wi, demonstrating that for smaller Wi, it takes longer to reach a specific, critical stress ratio of 1. The magnitude of this stress ratio can be interpreted as a structural parameter quantifying the recoverable strain or degree to which a fluid is "affine".    

\section{\label{sec:level1} Experimental Validity and Limitations}
Techniques for measuring transient extensional viscosity have limits on the maximum amplitude (Hencky strain) and the maximum Hencky strain rate. 

For example, uniaxial capillary breakup extensional rheometry and drip-on-substrate techniques (CaBER and DOS, respectively) are limited a specific Hencky strain rate (dependent on the relaxation time of the fluid)\cite{entov1997effect}, where Wi = 2/3. Additionally, the Hencky strain rate \textit{prior to} the elasto-capillary regime is driven by the balance between capillary and viscous forces, causing the effective Hencky strain rate to change (i.e. the flow is not a steady extensional flow at early strains). Commercial SER devices can only reach Hencky strain amplitudes of 4, but up to Hencky strain \textit{rates}\cite{tainstruments_evf2} of $10$ s$^{-1}$. 

In the case of exponential shear, the equivalent limitations that drive the maximum Hencky strain and strain rate arise from three sources: the maximum speed and ramp rate of the motor (section \ref{sec:MotorLimits}), the noise-floor and saturation limits of the transducers (for both the normal stress and the shear stress sensors; section \ref{sec:transducerLimits}), and limitations due to the generation of secondary flows and instabilities at high rates (\ref{sec:instabilities}). For several of these quantities, the limits on shear rate and stress limits become geometry-dependent. Here we address limitations associated with a TA-instruments ARES G2 rheometer with cone and plate versus parallel plate attachments. 

\subsection{Motor Limitations}
\label{sec:MotorLimits}
The maximum rotational velocity of the motor for the TA instruments Ares G2 is reported at approximately 320 rad/sec. In practice, we can set an arbitrary wave exponential shear protocol with a data rate of 1000 pts/sec, and the practically achievable rate is closer to 160 rad/sec (as the instrument stops if it \textit{predicts} it will go over a specific value of $\omega$. At this point, the instrument reports an error and requires software reset. The authors generally recommend staying under $160$ rad/sec if using the current (2025) software in arbitrary-wave mode. 

This limit of $\omega_{max}$ can be translated to a maximum shear rate on a cone and plate rheometer, given by: 
\begin{equation}
   \dot{\gamma}_{max}  = \frac{\omega_{max}}{\beta},
\end{equation}

and for a parallel plate is given by: 
\begin{equation}
       \dot{\gamma}_{max}  = \frac{\omega_{max} R} {h}.
\end{equation}

\begin{figure}[ht!]
\includegraphics[width =0.8\linewidth]{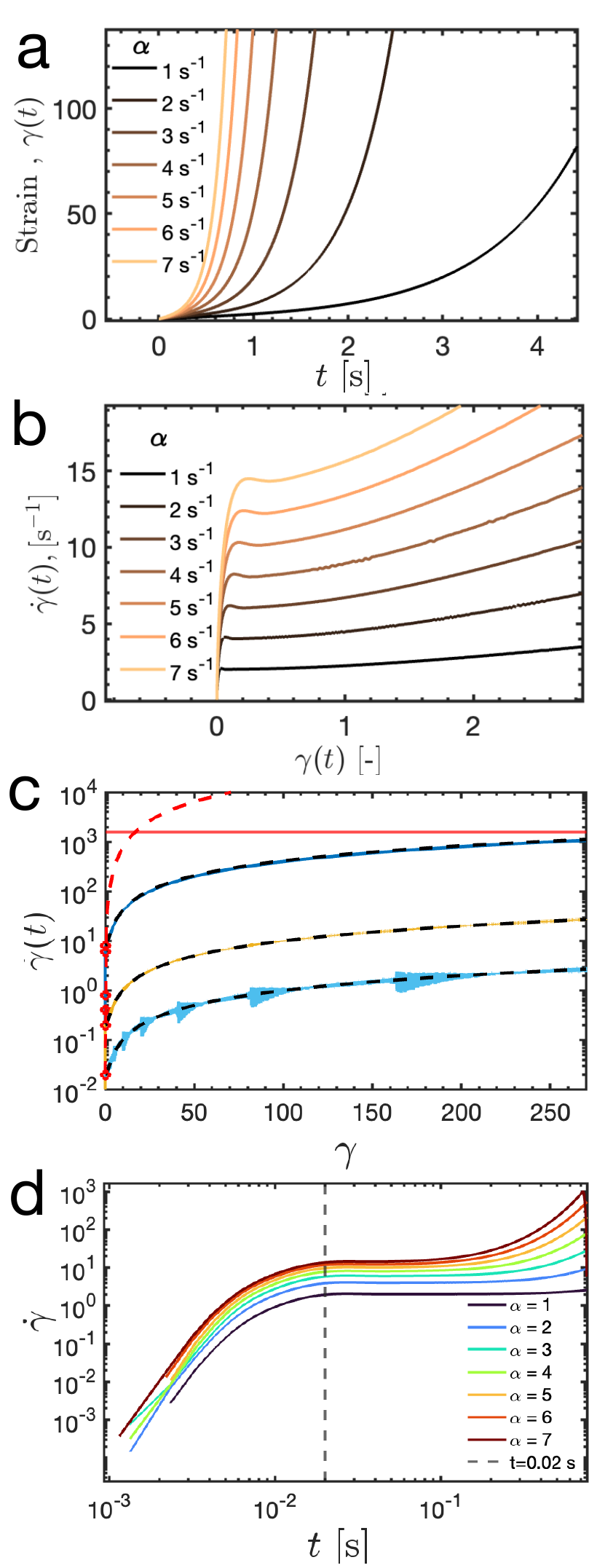}
\caption{\label{fig:6} Kinematic limitations of exponential shear arising from motor limits. (a,b) The actual transient strain waveforms for exponential shear, as produced on an ARES G2 rheometer are depicted in panels a and b, showing the strain and strain rate respectively. (c) The strain rate versus strain with the kinematic limits for a $25$mm diameter cone and plate attachment ($\beta= 0.1$ rad). The black dotted lines are the input waveforms for $\alpha = 0.01, .1 $ and $4$, where the solid colored lines are the actual positions of the instrument. Red lines show the kinematic limits discussed in section \ref{sec:transducerLimits}. (d) shows that the startup time of the instrument is approximately 0.02 sec}
\end{figure}

This maximum motor limit is shown in figure \ref{fig:6}c as the solid red line, for a cone and plate geometry ($d = 25$ mm, $\beta =0.1$ rad). 

In addition to a maximum speed of the motor, there is also a start-up time of the motor. Due to the prescribed strain rate of $\dot{\gamma}_{yx} = 2 \alpha \cosh(\alpha t)$, at t=0 the prescribed starting shear rate is $2 \alpha$. The Ares G2 has a start up time of approximately 0.02s, as shown in figure \ref{fig:6}d. In practice, this limit is not very constrictive, as shown in figure \ref{fig:6}c at the dotted red line. 

\subsection{Transducer Limits}
\label{sec:transducerLimits}

The normal stress transducer and the shear stress transducer also have respective limitations / maximum readable values. 

For a cone and plate device, the sensor limits are given by:
\begin{equation}
    \sigma_{\mathrm{min}} = \frac{3 T_{\mathrm{min}} }{2 \pi R^3},
\end{equation}

\begin{equation}
    N_{1,\mathrm{max, cp}} = \frac{F_{\mathrm{min}}}{(\frac{\pi R^2}{2})}.
\end{equation}

For parallel plate, the limits can be computed as:
\begin{equation}
    \sigma_{\mathrm{min}} = \frac{2 T_{\mathrm{min}} }{\pi R^3},
\end{equation}

\begin{equation}
    N_{1,\mathrm{max, pp}} = \frac{F_{\mathrm{max}}}{(\pi R^2)}.
\end{equation}

For the ARES G2, the stated detectable normal force range is $0.001$ to $20$ N, and the stated torque minimum is $10^{-7} \mathrm{N} \cdot \mathrm{m}$ .

\begin{figure}[ht!]
\includegraphics[width =0.9\linewidth]{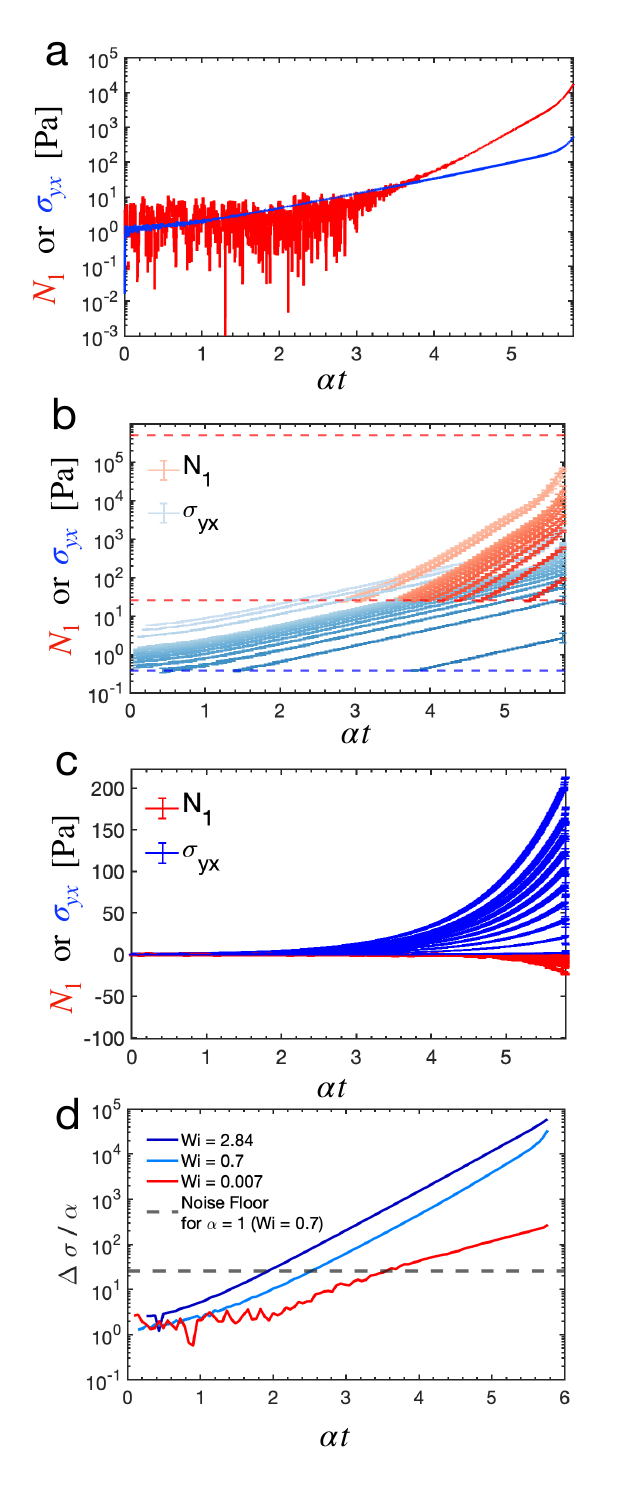}
\caption{\label{fig:sensorLimits} Sensor and Fluid Limitations (ARES G2 rheometer). (a) Raw data for the first normal stress difference (in red) and the shear stress (in blue) for a $0.3$\% wt PIB Boger fluid at $\alpha =2$. Note the noise, especially in the normal force data. (b) Filtered data, showing the normal stress sendor limits (red dotted) and shear stress minimum limit. (c) Newtonian fluid showing inertial contribution to $N_1$. (d) Computed noise floor for the total principal stress based on the equations in \ref{sec:transducerLimits} shows that the data appears above the formal noise floor only above Hencky strains of about $1.5-3$ in this experiment. This is the high-Wi "model-independent" material function discussed in section \ref{sec:modelInd}}
\end{figure}

\subsection{Limits from onset of edge fracture and fluid instabilities/secondary flows}
\label{sec:instabilities}
At very high strain rate, various kinds of fluid instabilities and edge fracture can cut-off the valid regime of data (even if the motor and sensors are within their operational limits. We refer the reader to section 3.3.2 of \textcite{ewoldt2015experimental} for rough approximations of the critical Reynolds numbers for cone and plate geometries. In practice, looking for a sudden increase in the shear stress at high strain amplitudes (often seen above $\alpha t = 5.8$ in polymeric fluids) is a strong sign of onset of an instability. Practically speaking, data should be cropped if such a sudden (almost vertical) increase in shear stress is observed.


\subsection{A simple material function (for high-Wi only)} 
\label{sec:modelInd}
The \textit{simplest} material function that is still useful in exponential shear is equation \ref{eq:Kwan}, recommended by \textcite{kwan2001experimental}: $\eta^+_{ES} = \Delta\sigma/\alpha$. This material function only approximates planar extension in the limit of affine, nonlinear elastic materials at high Wi, but requires no complex corrective terms. In figure \ref{fig:sensorLimits}d, we show how this material function breaks down for a 0.3\% wt PIB Boger fluid at intermediate and low Wi. Note that at low Wi (shown in red), the material function \textit{still} shows robust stress growth (even above the sensor limits of the instrument), when it should be at a steady state. We caution the user that this material function is \textit{not representative} of planar extension unless the Wi is $\gg$ 1 -- a limit we can approximate as "affine", as defined in section \ref{sec:recoverableStrain}. 

This is an appropriate material function for high Wi applications where the extension rate is extreme (i.e. testing fluids for rheological properties for fiber spinning applications, where Hencky strain rates can commonly exceed $5$ s$^{-1}$) -- \textit{or} if the sample is effectively a crosslinked Neo-Hookean elastic solid (for example, a slightly melted cheese or an elastic hydrogel). Both these examples would both fall in this simple "affine"/ high-Wi limit, as $\alpha \tau \gg1$.

The shortcomings of this simple material function however are also important: specifically, in examples where the user is trying to determine critical processing conditions, where the user has no knowledge of the rough relaxation time of the fluid, or when the user needs to test a viscoelastic fluid for properties at a specific, moderate Wi (all common situations in polymer processing, food science, pharmaceutical manufacturing -- among other industries).    

\section{\label{sec:unifying} A unifying, frame-invariant material function \protect\\ }

From equation 6, we can pose the question directly to compute the effective strain rate in exponential shear (the stretching rate \textit{necessary} to mimic a specific, constant Hencky strain rate in planar extension, $\dot{\varepsilon}$):

\begin{equation}
\label{eq:matFunc}
    \eta_{ES}^+ = \frac{\Delta\sigma_{\mathrm{experiment}}}{RF} = \frac{\Delta \sigma_{PE}}{\dot{\varepsilon}}
\end{equation}

Using equation 9 as a "ground truth" for the total principal stress in planar extension ($\Delta \sigma_{PE}$), we can derive what $RF$ needs to be to match the exponential shear material function to planar extension. Note that $\Delta\sigma_{\mathrm{experiment}}$ is given directly from experimental data (and may deviate from $\Delta \sigma_{ES}$) -- as both $N_1$ and $\sigma_{yx}$ are independently, simultaneously accessible from modern strain controlled rheometers <cite TA instruments>. 

This framing effectively allows for the effective stretching rate (RF) to be computed using a frame-invariant and physics-based method -- but does not pin the material function itself into a model. Non-ideal/ non-UCM fluid responses are still captured by experimental data in the numerator of the material function. 

After some non-trivial refactoring and term grouping, we find the UCM-derived rate factor (RF) to be analytically-tractable and of the form: 

\begin{align*}
\mathrm{RF} = {} & \frac{2 \alpha \left(4 \alpha^2 \btau^2 - 1\right)}
{\left| \alpha^2 \btau^2 - 1 \right| \left(\dfrac{\dot{\gamma}^2}{\alpha^2} + 4 \btau \gamma \dot{\gamma} + \gamma^2 - 4 e^{t/\btau}\right)}\\[1ex]
&\times \Biggl\{ \frac{4}{\left(1-4 \alpha^2 \btau^2\right)^2} \Biggl[
-\alpha \btau \Bigl(\left(4 \alpha^2 \btau^2 - 1\right) e^{t/\btau} - 2 \alpha^2 \btau^2 \\
&\quad + \frac{3}{2} \btau \gamma \dot{\gamma} e^{t/\btau} + 2 \Bigr)
+ \frac{\btau \left(2 \alpha^2 \btau^2 + 1\right) e^{t/\btau} \left(\alpha^2 \gamma^2 + \dot{\gamma}^2\right)}
{4 \alpha} \\
&\quad + \left(4 \alpha^2 \btau^2 - 1\right) \gamma \Biggr]^2 
+ \frac{\left(e^{t/\btau} \left(\alpha^2 \btau \gamma - \dot{\gamma}\right) + 2 \alpha\right)^2}{\alpha^2} \Biggr\}^{\frac{1}{2}}\,
\end{align*}

\vspace{-5 pt}
\begin{equation}
\label{eqn:RF}
    \mathrm{  ~~}
\end{equation}

Note that this rate factor (RF) does not include any instances of \( \eta_p \), as these terms cancel. 

\begin{figure}
\includegraphics[width= 0.9\linewidth]{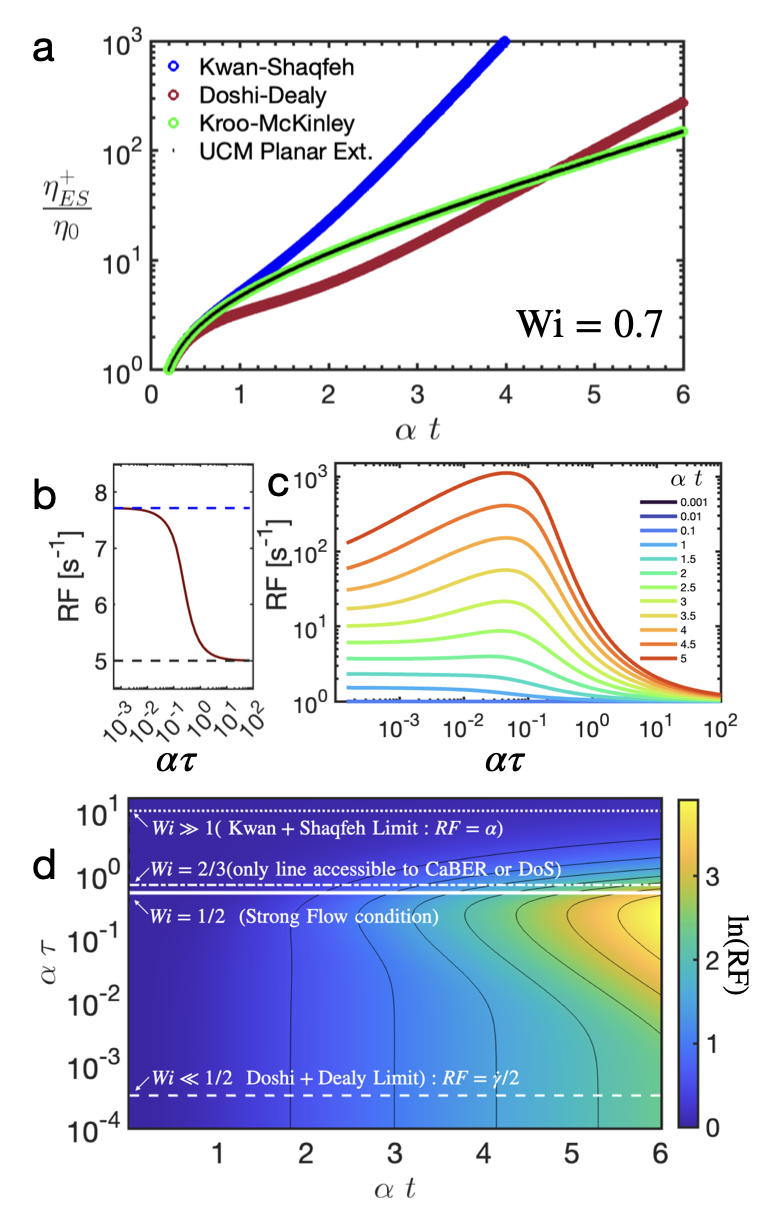}
\caption{\label{fig:RateFactor} a) Comparison between proposed material functions at Wi$ = 0.7$ for a synthetic UCM fluid with $\tau = 0.7$ and $\alpha = 1$. Note that our new proposed function is shown in green. Planar extension is shown in black. b)The rate factor (eq \ref{eqn:RF}) is a function of Wi. This is the RF plotted for different Wi at $\alpha t= 1$ and $\alpha = 5$, interpolating smoothly between the deformation rates previously proposed in literature ($\alpha$ versus $2 \dot{\gamma}$) . c) The rate factor also adjusts as a function of Hencky strain ($\alpha t$). Here we have shown this plotted for $\alpha =1$. Note that for some parameter combinations at high strain amplitudes, the behavior can appear non-monotonic as the Weissenberg number is varied. d) The dual dependence of the rate factor on $t$ and $\tau$ creates a phase space -- populating a Pipkin diagram with the effective deformation rate indicated on the colormap. The appropriate rate is both strain amplitude and strain rate dependent. In the limit of strong flows (Wi $> 0.5$), the RF is well approximated by the constant, $\alpha$. In the low Wi limit, the rate is well approximated by $\dot{\gamma}/2$ (or $2 \dot{\gamma}$ depending on if you are defining the extensional viscosity as relative to $4 \eta_0$ or $\eta_0$ (we suggest the former approach in this paper)}. 
\end{figure}

Figure \ref{fig:RateFactor}a demonstrates the effectiveness of the Rate Factor (equation \ref{eqn:RF}) at mimicking planar extension with synthetic data (UCM) at an intermediate Wi. Unlike previously defined strain rates (Figure \ref{fig:4}b), this new function (shown in green) captures the intermediary dynamics of planar extension (shown in black). 

This new material function also has a Wi dependency (as we predicted it \textit{should} in section \ref{sec:recoverableStrain})- as shown in Figure \ref{fig:RateFactor}b and c. The function smoothly interpolates between the two different previously proposed material functions (shown in blue dotted lines in Figure \ref{fig:RateFactor}b, and as defined in equations\ref{eq:Kwan} and \ref{matFuncDoshi}). Curiously, at large Hencky strain amplitudes and intermediate Wi, equation \ref{eqn:RF} has a maxima, where there exists a non-monotonic transition between the low Wi and high Wi regimes.

This is perhaps best visualized in Figure \ref{fig:RateFactor}d, where the contour map shows the RF for a range of Hencky strain amplitudes ($\alpha t$) versus Wi ($\alpha \tau$). In this phase space, any material in exponential shear would traverse from left to right along a given horizontal line. Exponential shear would be a "strong flow" for any horizontal line above $\alpha \tau = 1/2$. Elasto-capillary based extensional rheometry methods would be mimicked in exponential shear by traversing the line at $Wi = 2/3$. In the high Wi limit, the RF simply becomes $\alpha$ at all strains. In the low Wi / Newtonian limit,  $\dot{\gamma}/2$ (not 2 $\dot{\gamma}$), if relative to $4 \eta_p$ rather than $\eta_p$ in the Newtonian limit (this choice is consistent with the way we defined $\eta_p$ initially in  equation \ref{eqn:UCM}, a decision in agreement with \textcite{larson2013constitutive} but deviating slightly from \textcite{bird1977dynamics} and \textcite{doshi1987exponential}). 

The only parameter in Equation \ref{eqn:RF} not directly available from the experimental inputs is the relaxation time ( $\btau$ ). Incorporating ( $\btau$ ) into a material function is problematic, because the experiment does not \textit{appear} to directly determine ($\btau$); its evaluation would apparently require fitting the fluid behavior to an Upper-Convected Maxwell model. Fitting an Upper-Convected Maxwell model universally to any complex fluid is inappropriate. Although the model is frame invariant and capable of capturing certain nonlinear behaviors, UCM is a simplistic representation that fails to quantitatively predict the diverse responses of viscoelastic fluids -- which typically exhibit a spectrum of relaxation times that each contribute to the nonlinear response. Nonetheless, the presence of ($\btau$) in the RF equation indicates that the effective rate of material stretch depends on a Weissenberg number, ( $\mathrm{Wi} = \alpha \btau $) - a complexity that requires additional insight. 

\subsection{\label{sec:tstar} Approximating $\btau$ from the Stress-Growth timescale, $t^\star$}

Taking a Buckingham-Pi-style approach, 
the approximate average relaxation time ($\btau$) represents a single timescale that is characteristic or inherent to the rheological response of the fluid. For the material function to be complete, we need this characteristic timescale to be accessible directly from exponential shear experiment data. 
\begin{figure*}
    \centering
    \includegraphics[width=0.8\linewidth]{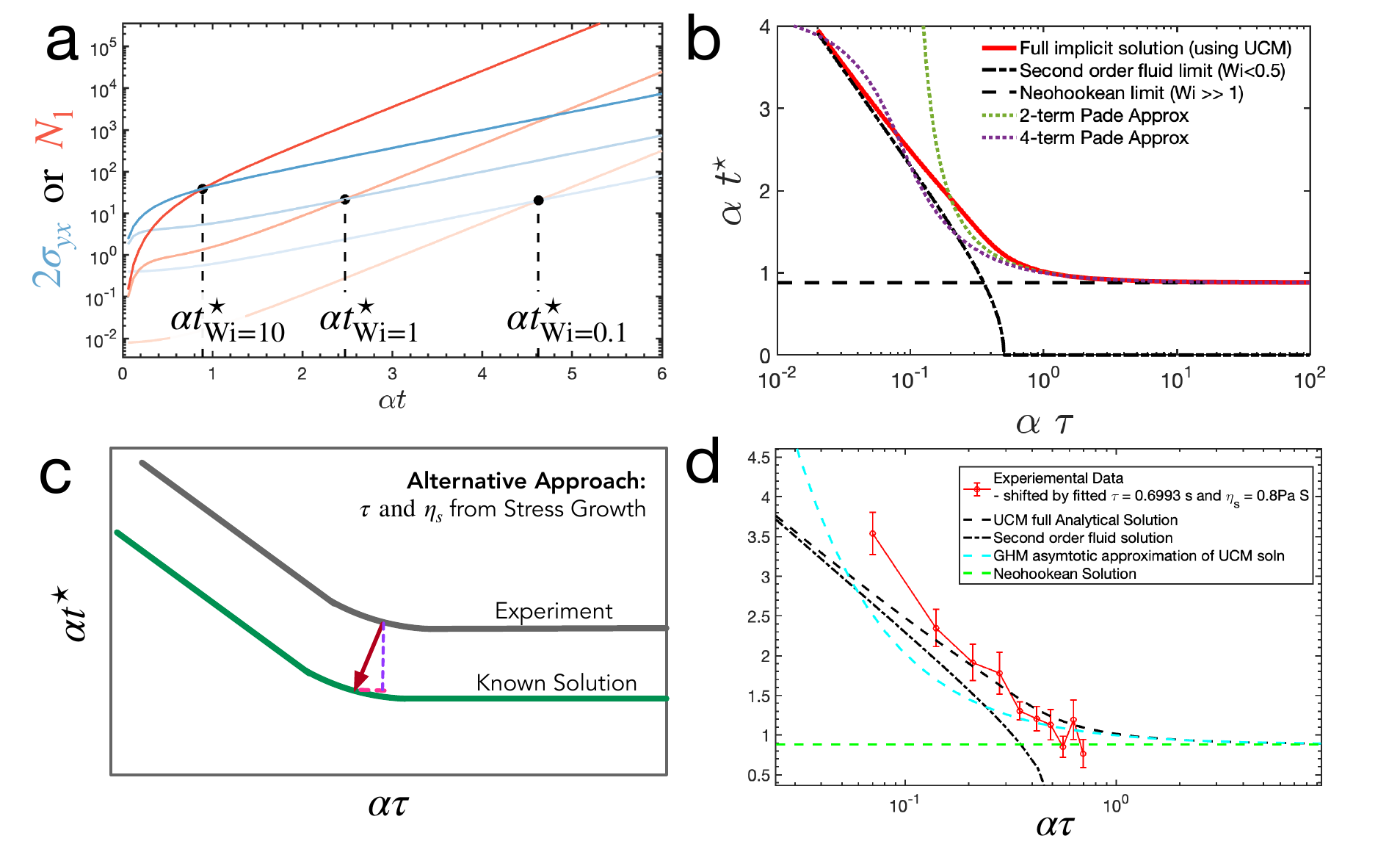}
    \caption{(a) There is a characteristic timescale associated with exponential shear that is shown in this plot. It has to do with when the normal stress and twice the shear stress become equal (or multiples) of each other. The time at which this occurs we call $t_\star$ (the stress growth timescale). (b) This timescale can be directly related to the relaxation time to infer an approximate Wi number. Here we show an Upper-Convected Maxwell fluid (red) interpolating between the two limits of a second-order-fluid and a Neo-Hookean Solid. (c) An approach to find the relaxation time (if the solvent viscosity is known is proposed. If neither the solvent viscosity nor the relaxation time are known, they can be fitted simultaneously if the user has performed exponential shear on the fluid at several different strain rates. (d) We demonstrate this for an example 0.3\% wt PIB solution.}
    \label{fig:tstarMethod}
\end{figure*}

Upon careful inspection, there indeed exists a characteristic timescale that is \textit{observable} from the relative stress growth of $N_1(t)$ versus $\sigma_{yx}(t) $. Specifically, we can define a \textit{different} characteristic time as the point where the stress ratio (from eq XYZ) is equal to 1: 
\begin{equation} 
    N_1(t^\star) = 2 \sigma_{yx}(t^\star)
    \label{eq:stressBalance}
\end{equation}

We then pose the question: does this (observable) stress-growth timescale imply an (approximate) relaxation time of the fluid that might be used to inform the material function? 

Specifically, can we approximate:  
\begin{equation}
    \btau \approx\tau_{exp} = f(t^\star, \alpha)
\end{equation}

where $\tau_{exp}$ is not the Upper-Convected Maxwell relaxation time, but an \textit{intrinsic} experimental relaxation time. 

Using the same idea from section 2 of using synthetic data from a UCM fluid we can plug in Eq \ref{eq:N1} and \ref{eqn:sigmayxUCM} to Eq \ref{eq:stressBalance}, and attempt to solve for the idealized form of this relationship. 

Unfortunately, this inversion leads to a transcendental equation ($\tau$ appears both inside the natural logarithm and outside of it) and can only be solved explicitly within specific limits. The function itself is strongly non-linear, relating a relaxation time that realistically could span many orders of magnitude to a practically observable parameter that spans less than one order ($0.88 <\alpha t^\star < 4 $). 

While this is simply how one example fluid (a UCM fluid) relates the stress-growth time scale and the approximate relaxation time -- we observe a number example fluids to examine the \textit{generality} of this approximation of the relationship between $\tau$ and $t^\star$. 

In the high Wi limit, the Neohookean solution is given by:
\begin{equation}
    \alpha t^\star = \mathrm{sinh}^{-1}(1)
\end{equation}

The second order fluid limit is given by: 

\begin{equation}
\label{eq:SOF}
\alpha t^\star = \cosh^{-1}(\frac{1}{2\alpha\tau})  
\end{equation}

Remarkably, the full UCM solution interpolates cleanly between these two limits, suggesting a general relationship for viscoelastic fluids relating an average relaxation time to an observed stress-growth timescale. 

Because the relationship between $\tau$ and $t^\star$ that satisfies equation~\eqref{eq:stressBalance} is transcendental - in practice, it is expediant to invert this algebraic equation numerically for $\tau(\alpha t^\star)$, using the implicit relationship (ie $\alpha t^\star = f(\tau)$).  

However, this is somewhat irritating for back-of-the-envelope estimations. We \textit{can} approximate the explicit solution using a Taylor series expansion in the Neo-hookean limit (i.e. fixed point at $asinh(1)$ ) to solve for an approximate solution.  

Proceeding with this approach, a 2-term approximate solution becomes:
\begin{equation}
\label{eq:taylorSeries}
\alpha t^\star \approx   \sinh^{-1}(1) \cdot \left( 1 + \frac{\sinh^{-1}(1)^2}{6 \, \alpha \, \btau} ... +O(3)\right)
\end{equation}

We can invert \ref{eq:taylorSeries} and reorganize this equation into an explicit, rational function (now also showing also the next term in the series) shown in green in figure \ref{fig:tstarMethod}. This Padé-type approximation can be written: 
\begin{equation}
\label{eq:tauSoln}
    \btau \approx \frac{\sinh^{-1}(1)^3}{6 \, \alpha \left( \alpha t^\star - \sinh^{-1}(1) \right)} 
           + \frac{\sinh^{-1}(1)}{10 \, \alpha}
\end{equation}

For two terms, we achieve good accuracy down to Wi around $0.7$. For more terms (say 4 terms), we begin to achieve approximate accuracy down to somewhat lower Weissenberg number (shown in purple in figure \ref{fig:tstarMethod}b). In practice, the authors recommend using the implicit UCM-based solution (which uses the actual frame-invariant UCM-inspired method to interpolate between the second-order fluid limit and the Neohookean limit), such that that no unnecessary errors are introduced through approximating this non-linear function. These analytical forms are particularly useful only for rough approximations of the relationship in the strong flow regime at Wi > 0.5. 

From a practical perspective, when $\alpha t^\star$ approaches $\mathrm{asinh}(1)$,  $\btau$ becomes indeterminate, since there is an asymptote. Tiny changes in $t^\star$ would lead to huge changes in $\btau$. This may appear to be a problem -- but when the data falls in this regime, the rate factor (eqn \ref{eqn:RF}) \textit{also} asymptotes to  $RF = \alpha$ -- This high- Wi "affine" limit removes the need for Eqn \ref{eqn:RF} entirely.

\subsection{\label{sec:cessation} Estimating $\tau$ from cessation of Exponential Shear}
\label{sec:cessation}
\begin{figure}
    \centering
    \includegraphics[width=\linewidth]{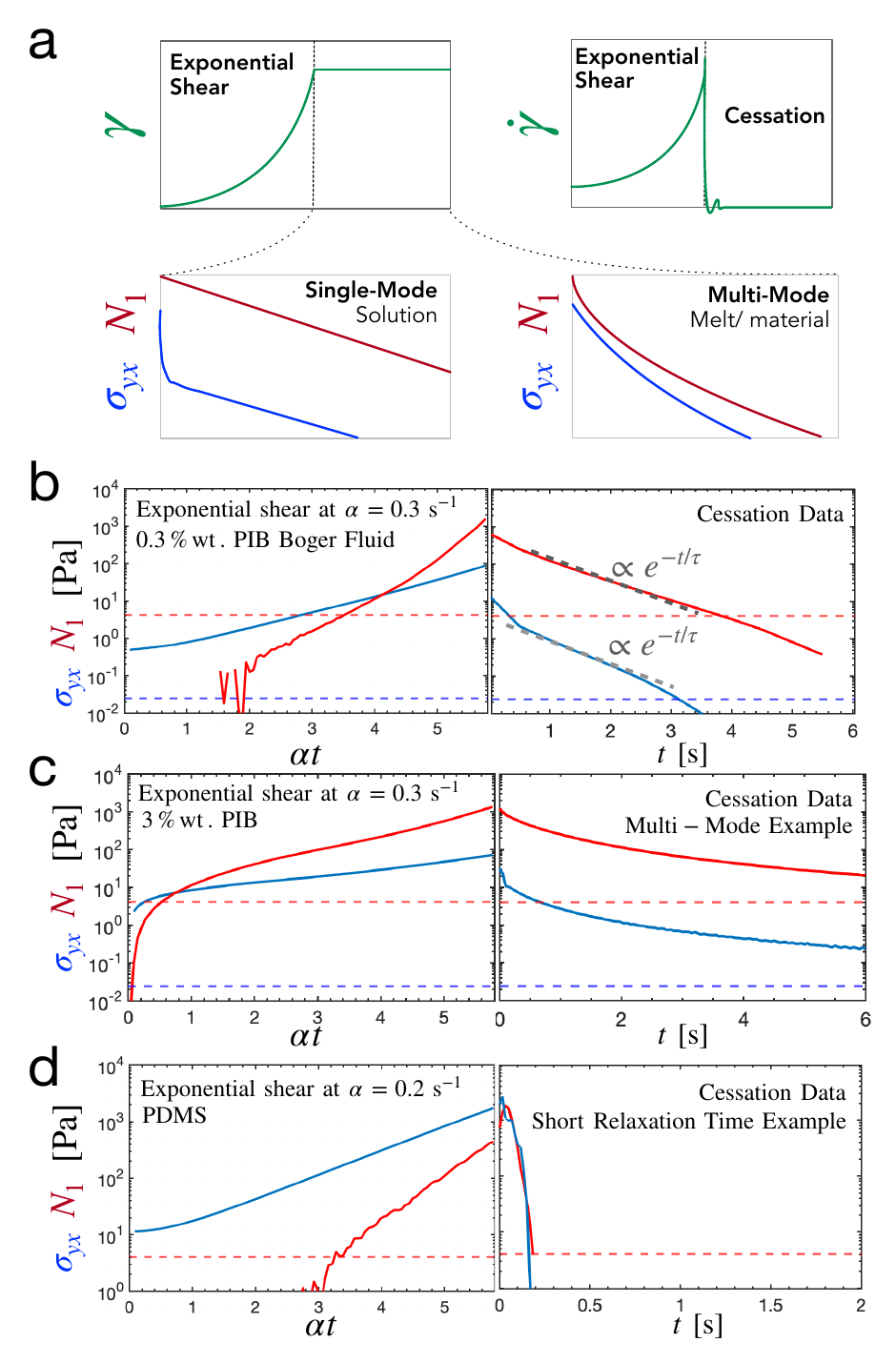}
    \caption{(a) The concept of recording the cessation of exponential shear allows for additional experimental observations. In most cases, $\tau$ and $\eta_s$ can both be extracted from the shear stress data upon cessation of exponential shear. (b)Example cessation of a PIB Boger fluid with a single, dominant relaxation time (c) Cessation of a multi-mode fluid complicates the extraction of $\tau$. Here we show a multi-mode shear-thinning $3\%$ wt. PIB solution. (d) very short relaxation times can be extracted (PDMS sample shown) in cases where the normal stresses are not sufficiently large to use the $t^\star$ method described in section \ref{sec:tstar}. 
    \label{fig:cessation}}
    
\end{figure}
An alternative method to estimate $\btau$ is to observe the relaxation of the fluid after cessation of exponential shear. This gives direct, experimental insight into this same approximated quantity, $\tau_{exp}$ as the $t^\star$ method in section \ref{sec:tstar}. The concept is depicted in figure \ref{fig:cessation}a: upon cessation of exponential shear, the shear stress and the normal stress decay. For small amplitudes and small $\alpha$ values, the decay in the shear stress contains the relaxation modes. 

Specifically, for simple viscoelastic fluids (like the $0.3\%$ wt PIB Boger fluid depicted in \ref{fig:cessation}b, the normal and shear stresses decay approximately in parallel, $\propto e^{-t/\tau}$. For larger amplitudes, ringing in the shear rate can cause errors in this measurement, especially for short-relaxation time fluids. Nonetheless, we have found the method to be reasonably robust at low to intermediate Hencky strain rates. 

Multi-mode fluids contain several relaxation times (as shown in figure \ref{fig:cessation}c), and are therefore more difficult to analyze. In principle, one could fit several modes and then weigh the modes to achieve a single value for $\tau$ from this data (as is commonly performed in traditional model-fitting, for example, in Appendix A). In this study however, we have simply chosen a range in time where there appears to be a dominant mode and extract the longest relaxation time from fitting to the slope (see table \ref{tab:table1}). For fluids that don't have a clear single-mode to extract, we have conservatively marked 'multi-mode' and defer to the alternative method in section \ref{sec:tstar}. 


\subsection{\label{sec:corrections} Necessary Corrections to $ N_1$ and $\sigma_{yx}$}
\label{sec:corrections}
\subsubsection{Solvent Viscosity}
Solvent viscosity is arguably the biggest complicating factor in exponential shear. Fundamentally, a perfectly Newtonian solvent should not contribute to the stress growth dynamics of the extensional viscosity at all. But, in shear it (inconveniently) adds a constant contribution to the total viscosity; i.e.: 
\begin{equation}
    \eta_0 = \eta_s + \eta_p
\end{equation}

This effectively modifies any measurement taken with an additional (and unwanted) time-dependent term in the (measured) shear stress:
\begin{equation}
    \sigma_{\mathrm{meas}}(t) = \eta_s \dot{\gamma}(t) + f(\eta_{p,\mathrm{meas}}(t))
\end{equation}

where $f(\eta_{p,\mathrm{meas}})$ in this case is the real, observed experimental behavior of the fluid (\textit{not necessarily} a function of similar form to equation \ref{eqn:sigmayxUCM}). 

Because the solvent stress has a shear-specific effect on the apparent viscosity, we need to subtract this quantity in the numerator of our material function (to compute the effective total principal stress in extension, considering only polymeric contributions). This same adjustment is also necessary prior to using the $t^\star$ method described in section \ref{sec:tstar}. This is a reasonable approach because in extensional flows, only polymeric stresses contribute to stress growth (i.e. $ \Delta \sigma_\mathrm{meas} = \dot{\varepsilon}\eta_p $). 

This is the primary distinguishing factor between the Upper-Convected Maxwell model (UCM) \cite{larson2013constitutive} and a single-mode Oldroyd-B model \cite{bird1977dynamics} (with a zero second normal stress coefficient). Indeed, using the Oldroyd-B model as an example, we can also observe that the normal stress difference in exponential shear remains completely unaffected by the modification of the solvent viscosity (because pure Newtonian solvents cannot generate normal stresses): 

\begin{equation}
     N_{1,\mathrm{meas}}(t) = f(\eta_p(t))
\end{equation}
   
If the solvent viscosity is unknown a-priori, a key challenge is how to determine $\eta_s$ robustly from exponential shear data directly. 

The simplest method is to observe the immediate drop in shear stress upon cessation of exponential shear. This is depicted in figure \ref{fig:cessation}b for a $0.3\%$ wt PIB Boger fluid. The product of the maximum shear rate at cessation and the observed instantaneous drop in stress is a good estimate of the shear stress:

\begin{equation}
  \eta_s =  \sigma_{yx, \mathrm{drop}}/ \dot{\gamma}_{max} 
\end{equation}

This method is quite robust, even for relatively modest solvent viscosity. We report solvent viscosity of several different fluids using this method in Table 1. One downside of this method is that it can be easy to conflate short-time relaxation modes with the drop due to solvent viscosity.  

The alternative method to find $\eta_s$ is to simultaneously optimize a best fit for both $\tau$ and $\eta_s$ from figure \ref{fig:tstarMethod}c  -- the experimental data can be shifted in 2 dimensions and we have 2 unknowns in the problem. The key however is that the exponential shear data must contain multiple $\alpha$ values for this method to work robustly (which is not ideal). Although it  does appear to estimate $\eta_s$ reasonably well (see: table \ref{tab:table1}) (within an order of magnitude), we have found it preferable to estimate this from the cessation data (figure \ref{fig:cessation}) where possible.   

\subsubsection{Fluid Inertia}
Similar to how there are shear-specific effects that influence the shear stress, there is an important shear-specific effect that also influences the first normal stress difference at high rates: inertia. \textcite{macosko1994rheology} gives a simple form to correct the first normal stress from contributions due to inertia:

\begin{equation}
    N_1 = N_{1,\mathrm{meas}} + 0.015 \cdot\rho\cdot\omega^2
\end{equation}

where $\rho$ is the density of the fluid and $\omega$ is the rotation rate (for cone and plate given by $\omega = \dot{\gamma}_{yx}\beta $). Similar to the process of correcting the shear stress, this correction should be made to all $N_1$ data -- correcting the total principal stress (the numerator of our material function \ref{eq:matFunc}), and also applying a correction to the method described in section \ref{sec:tstar}, to identify $ \btau$ from $t^\star$.  

\subsection{\label{sec:badData} "How to Avoid Bad Data" in exponential shear}
Exponential shear by nature is an advanced rheological method - more nonlinear and aggressive than any other technique the authors are aware exists in the community presently for strain-controlled rheometers. Compared to LAOS, amplitude sweeps, or flow curves, exponential shear can (transiently) achieve shear rates on samples that would onset instabilities and inhomogeneities in steady or periodic flows at equivalent rates. In this section we highlight a few practical experimental tips (in the spirit of \textcite{ewoldt2015experimental}).  

An intuitive picture of exponential shear might be to describe it as a fancy, high-amplitude step-strain-like protocol. The protocol similarities are that: (1) exponential shear is often very short in duration($ < 1 $s), (2) can reach a high strain amplitude, and (3) the cessation/ stopping of the exponential allows the user to also interpret the relaxation dynamics, as shown in Figure \ref{fig:cessation} (just as is done in step-strain or cessation of steady shear). The key difference is that a fast, aggressive prescribed strain ramp is used instead of a true step -- and data measurements (at least at 1000 data points/ sec) for $N_1$ and $\sigma_{yx}$ are collected \textit{during} the ramp itself.

Exponential shear as a technique (especially for low zero-shear viscosity fluids) is highly sensitive to sample trimming / test region filling. A poor contact line  on any part of the edge of a sample can quickly lead to discrepancies in the data depending on if the user is performing the test clockwise versus counterclockwise. For these reasons, we advise always using a sterile blade for trimming the sample (i.e. disposable razor blade), and cleaning the region on the plate surrounding the sample with an immiscible solvent. For example, for our oil-based polymer solutions, cleaning the bottom plate with acetone and water right up to the visible contact line reduces creep of the line at high shear rates. 

We also advise repeating the experiment multiple times, specifically averaging of at least 3 trials, prior to reporting an extensional viscosity. A standard deviation should always be reported with the stress data. Every line pictured in every panel of Figures 11-14 is the average of 4 separate data runs (2 clockwise and 2 counterclockwise) and we visualize error bars (Figures 11a and 12a), prior to running the material function analysis. This should be considered not-optional for this specific technique, since it is prone to user-error issues with precision sample mounting, and somewhat unpredictable onset of instability at higher strains. Unlike other large-amplitude oscillatory or steady-state measurements (which typically will average rheological properties over some integer number of cycles or some duration, respectively), exponential shear is inconveniently transient in nature. This is especially the case when pushing towards the upper limits of Hencky strain amplitude (i.e. $\varepsilon \approx 6$) with geometries that will further complicate the flow (as we do in this investigation in Figure \ref{fig:finiteExtensibility}), such as parallel plate. In such cases, we advise taking highspeed images of the sample to ensure the sample does not eject over the period of data collection. 

As sketched in Figure \ref{fig:cessation}a, the rapid deceleration from the maximum strain rate to zero at large prescribed $\alpha$ values can cause backlash in the instrument. At the highest rates, (aside from imaging the sample area to ensure the sample does not prematurely eject), consider mounting a new sample for each run, when operating at more aggressive rates (as we have done in Figure \ref{fig:finiteExtensibility}).  

\section{\label{sec:proof} Proof of Principle: testing the method on different fluids \protect\\ }

To demonstrate proof-of-principle we compute the transient extensional viscosity using this new proposed material function, as defined by equations \ref{eq:matFunc}, \ref{eqn:RF}, and sections \ref{sec:tstar} and\ref{sec:cessation}. To do this, exponential shear experimental data was used for four different test fluids, with no fitting parameters and no additional experimental data (i.e. no frequency sweeps, etc.). The transient extensional viscosity was computed versus $\alpha t$ (Hencky strain) for a range of different Wi for each fluid. Appropriate corrections are applied to the measured $N_1$ and $\sigma_{yx}$ values, as described in detail in section \ref{sec:corrections}. Importantly, we demonstrate the accuracy and limitations of this method on a diversity of samples that are \textit{not} all model, single-mode viscoelastic fluids. For each of these four fluids, the tests were all completed at 25 deg C, with a 25 mm cone and plate attachment ($\beta$ =0.1 rad), for values of $\alpha$ ranging from 0.01 to 4. 

A brief summary of key the experiential parameters ($\tau$ and $\eta_s$) computed using the analysis described in sections \ref{sec:tstar} and \ref{sec:cessation} are included in table \ref{tab:table1} (in blue text). For validation purposes, we also have included a comparison with Rouse-Zimm \cite{larson2013constitutive} and Multi-Mode Oldroyd-B\cite{bird1977dynamics} model-based fits (in red text). Details on standard fitting methods are in Appendix A. The table demonstrates that both approaches presented in section \ref{sec:tstar} and\ref{sec:cessation} are effective at determining these key parameters from exponential shear data.

\begin{table*}[!ht]
\setlength{\extrarowheight}{4pt}

    \centering
    \small
    \caption{Comparison between \textcolor{blue}{experiment-intrinsic parameters} (extracted directly from exponential shear data) versus \textcolor{red}{classically-fitted parameters} (numerically fitted to two common models from a small-amplitude oscillatory shear frequency sweep). In principle, the estimates of $\textcolor{blue}{\tau}$ and $\textcolor{blue}{\eta_s}$ (that are completely intrinsic to the experiment) are the only parameters needed to make the RF (eqn \ref{eqn:RF}) model-independent. Values marked with footnote$^1$ : $\tau$ is too small to compute because experimental data below instrument limits. Value marked N/A$^2$: Means the model did not converge on a solution. In this case, fluid $c\gg c^\star$; Rouse-Zimm not appropriate for highly entangled polymer solutions. N/A$^3$: Time resolution of averaged cessation data insufficient to capture this small of a solvent viscosity. Last 2 columns denote root mean squared error for model fits.  }
    
    \label{tab:table1}
    \begin{tabular}{cl|llll|lll|ll}
    &&Relaxation time&&&& Solvent Viscosity&&& Error \\
    \toprule
     & \textbf{Fluid} & \textbf{$\textcolor{red}{\bar{\tau}}_{\textcolor{red}{\mathrm{OLD-B}}}$} & \textbf{$\textcolor{red}{\tau}_{\textcolor{red}{\mathrm{RZ}}}$} & \textbf{$\textcolor{blue}{\tau_{f(t^\star)}}$} & \textbf{$\textcolor{blue}{\tau_{\mathrm{cessation}}}$} & \textbf{$\textcolor{blue}{\eta_{s,f(t^\star)}}$} & \textbf{$\textcolor{blue}{\eta_{s,\mathrm{cessation}}}$} & \textbf{$\textcolor{red} {\eta_{s,\textcolor{red}{\mathrm{OLD-B}}}}$} & \textbf{$\textcolor{black}{\mathrm{RMSE}}_{\textcolor{black}{\mathrm{OLD-B}}}$} & \textbf{$\textcolor{black}{\mathrm{RMSE}}_{\textcolor{black}{\mathrm{RZ}}}$} \\
    \toprule

    $1$ & Silicone Oil              & N/A$^{1}$   & N/A$^{1}$   & N/A$^{1}$   & N/A$^{1}$   & N/A$^{1}$   & --   & 1.01 Pa s   & N/A$^{1}$ & N/A$^{1}$ \\
    $2$ & PDMS                              & $0.786 \times 10^{-2}$ s   & $1.02 \times 10^{-2}$ s   & $0.59 \times 10^{-2}$ s  &multi-mode   & 6.84 Pa s  & 4.79 Pa s   & $5.67$ Pa s & 11.2 Pa & 15 Pa \\
    $3$ & 0.3\%wt PIB BF                   & 0.76 s & 0.773   & 0.699   & 0.709   & 0.80 Pa s   & 0.44 Pa s   & 0.601 Pa s & 0.0294 Pa & 0.111 Pa \\
    $4$ & 3\% wt PIB Fluid          & 1.69 & N/A$^2$   & 1.9647   & 1.6394 s   & 0.71 Pa s   &  N/A$^3$  & 0.238 Pa s   & 0.26 Pa & N/A$^2$  \\
    
    \end{tabular}
    
\end{table*}

\subsection{\label{sec:oil} UA2400 Viscosity-Standard Silicone Oil}
\begin{figure}

    \centering
    \includegraphics[width=\linewidth]{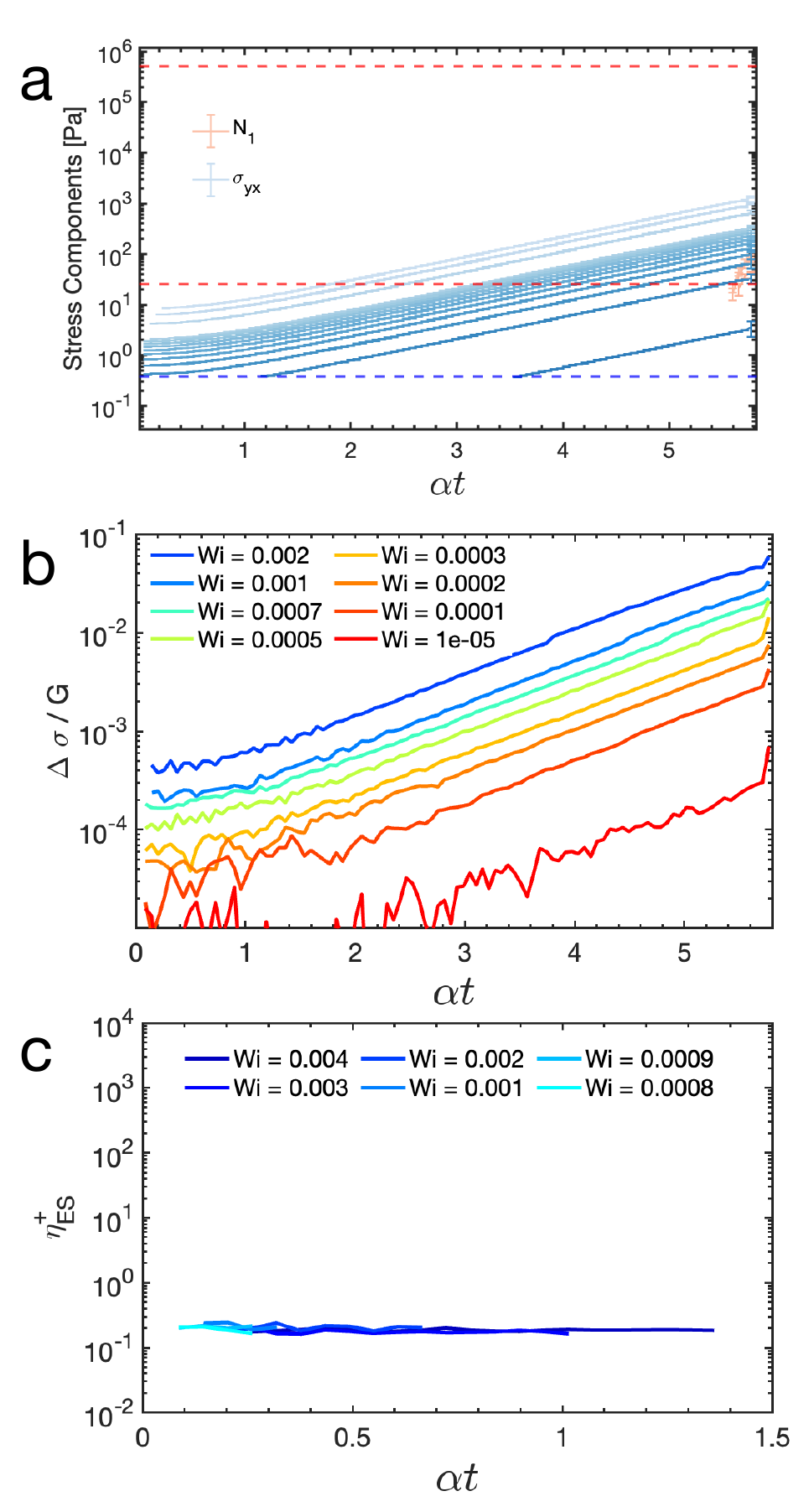}
    \caption{\label{fig:siliconeOil}Viscosity-Standard Silicone Oil. (a) Shear stress and normal stress components, with limits in dashed lines. (b) Total principal stress (c)Transient extensional viscosity shows no extensional thickening for a Newtonian fluid.}
\end{figure}

Starting with a 1 Pa-s viscosity-standard silicone oil, we demonstrate the output of this material function when there is a vanishingly small elastic contribution. As shown in figure \ref{fig:siliconeOil}a, the shear stress is measurable but effectively no normal stress difference is measured at any of the accessible values of Hencky strain rate ($\alpha = 0.1$ s$^{-1}$ to $4$s). 

Despite this, the normalized total principal stress in figure \ref{fig:siliconeOil}b still appears to exhibit stress growth. We assume a very small relaxation time (since it is so small it is immeasurable) of $\tau =10^{-4}$. Even at Wi$\sim O(10^{-5})$, it appears that stress is accumulating, highlighting the importance of an appropriate material function . 

If we use this experimental data with our material function from section \ref{sec:unifying}, the transient extensional viscosity, plotted in figure \ref{fig:siliconeOil}c vanishes for small Wi, and some noise at the slightly higher Wi appears to suggest the fluid is reporting a steady state (as expected by Trouton's ratio at steady state).

\subsection{\label{sec:pdms} PDMS}
\begin{figure}

    \centering
    \includegraphics[width=\linewidth]{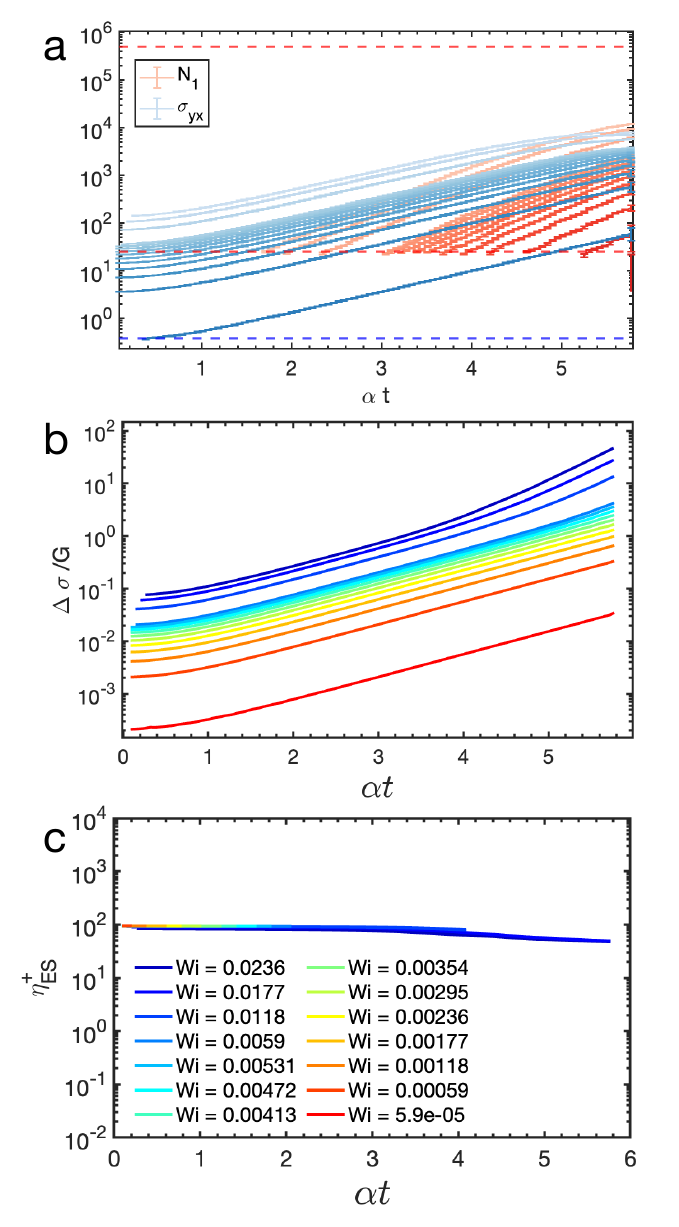}
    \caption{\label{fig:PDMS}PDMS/representative second-order-fluid-like sample (a) Shear stress and normal stress components, with limits in dashed lines. (b) Total principal stress (c)Transient extensional viscosity using$ ~\eta_s = 4.79$ Pa s and $\tau = 0.59 \cdot 10^{-2} s$ from experimental data in table \ref{tab:table1}. Although second normal stresses are accurately measured (within the valid regime for the transducer), the material function is able to accurately capture that this second order fluid does not experience extensional thickening for any of the Hencky strains tested.   }
 
\end{figure}

A similar approach is taken with a polydimethylsiloxane sample (PDMS): a silicone-based polymer melt. 

Unlike in the case of the viscosity-standard oil, the normal stress difference is measurable. However, $N_1$ only exceeds the  shear stress for a few of the largest values of $\alpha$ (specifically $ \alpha = 2,3,4 $s$^{-1}$). Using the method in section \ref{sec:tstar}, the relaxation time  $\tau_{f(t^\star)}$ was computed from these specific trials. This was used to process the material function data shown in figure \ref{fig:PDMS}c. 

Although the normal stresses are measurable, none of the accessible values of $\alpha$ are sufficient to reach Wi = 0.5. Similar to the viscosity-standard oil, the extensional viscosity is effectively constant, and collapses to a single steady-state "weak flow" where vorticity dominates. This is an example of a second-order fluid (quasi-linear) -- where a first normal stress is finite and present, but the fluid has a short enough relaxation time that it cannot realistically be driven above Wi = 0.5. This second order fluid limit is described in equation \ref{eq:SOF}, and represented as the black dot-dash line in figure \ref{fig:tstarMethod}b.  

\subsection{\label{sec:pibboger} $0.3\%$ wt PIB Boger Fluid}
\label{sec:PIBFluid}
\begin{figure}
    \centering
    \includegraphics[width=\linewidth]{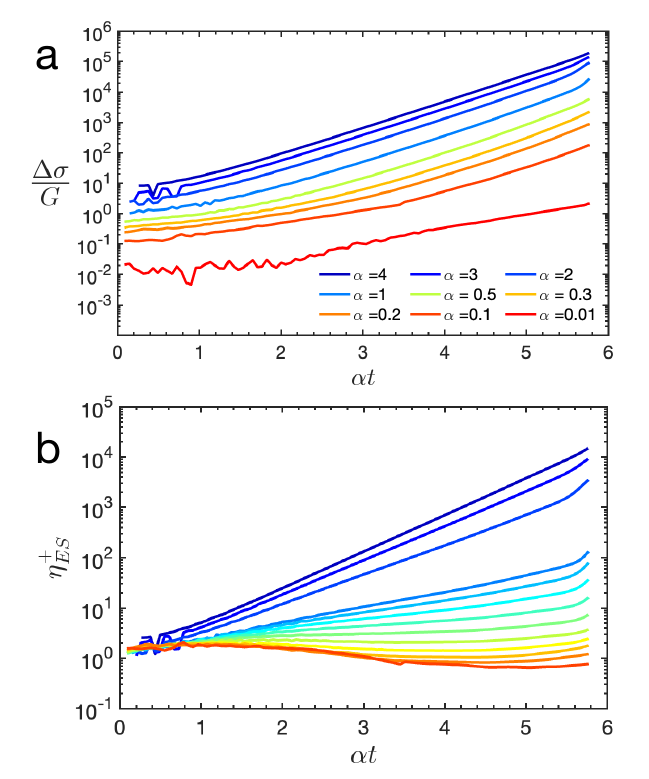}
    \caption{\label{fig:BF} $0.3 \% $ wt PIB Boger Fluid: a nonlinear viscoelastic fluid  (a) Normalized total principal stress (c)Transient extensional viscosity using$ \eta_s = 0.44$ Pa s and $\tau = 0.709$s from experimental data in table \ref{tab:table1}}
    \label{fig:enter-label}
\end{figure}

In figure \ref{fig:BF}, the nonlinear case of a $0.3\%$ wt polyisobutylene (PIB) fluid in exponential shear is shown (the same material used in this recent rod-climbing study \cite{more2023rod}). The molecular weight is approximately $10^6$ g/mol, dissolved in a mixture of a paraffinic oil-based solvent and a viscous mineral oil (to create a Bogerized version of the fluid\cite{boger1977demonstration}). The concentration of PIB in this solution is estimated to be close to the reported overlap concentration in \textcite{more2023rod} of $c^* =0.23 \%$ wt. 

Although the normalized total principal stress appears to rise for any value of $\alpha$ in the range of $0.01$ to $4$ s$^{-1}$, after the new material function is applied (figure \ref{fig:BF}b), the classic transition from a steady state to a stress growth is observed. The values of $\eta_{s,\mathrm{cessation}} = 0.44$ and $\tau_{f(t^\star)} = 0.709$ were used from table \ref{tab:table1}. We note that this is an excellent example fluid to demonstrate either the cessation-method (sec \ref{sec:cessation}) or the $t^\star$ method (sec \ref{sec:tstar}) -- both consistently are able to identify the parameters on interest ($\eta_s$ and $\tau$).   

\subsection{\label{sec:shearThinning} Shear Thinning 3\% wt PIB Solution}
Finally, we examine a highly concentrated solution containing $3\%$ wt polyisobutylene (PIB), dissolved in a mixture of more viscous paraffinic oil-based solvents. Via standard rheological characterization (see: Appendix A), it is clear that this fluid is shear-thinning (dropping from about 4.9 Pa s to 0.2 Pa s over roughly 3.5 decades of frequency, in a small amplitude oscillatory shear sweep).  

The rate factor analysis method was applied on this fluid: a nonlinear viscoelastic fluid that is neither a single mode, nor similar to an Upper-Convected Maxwell-type fluid. We find that the method \textit{still is able to capture the transition} from weak to strong flow at approximately a Wi = 0.5. At high Wi, (such as Wi $= 5.89 $and$ 7.86$), we observe a saturation in the stress growth (similar to the Neohookean saturation limit we expected from figure \ref{fig:UCMresult}b. 

\begin{figure}[ht!]

    \centering
    \includegraphics[width=1\linewidth]{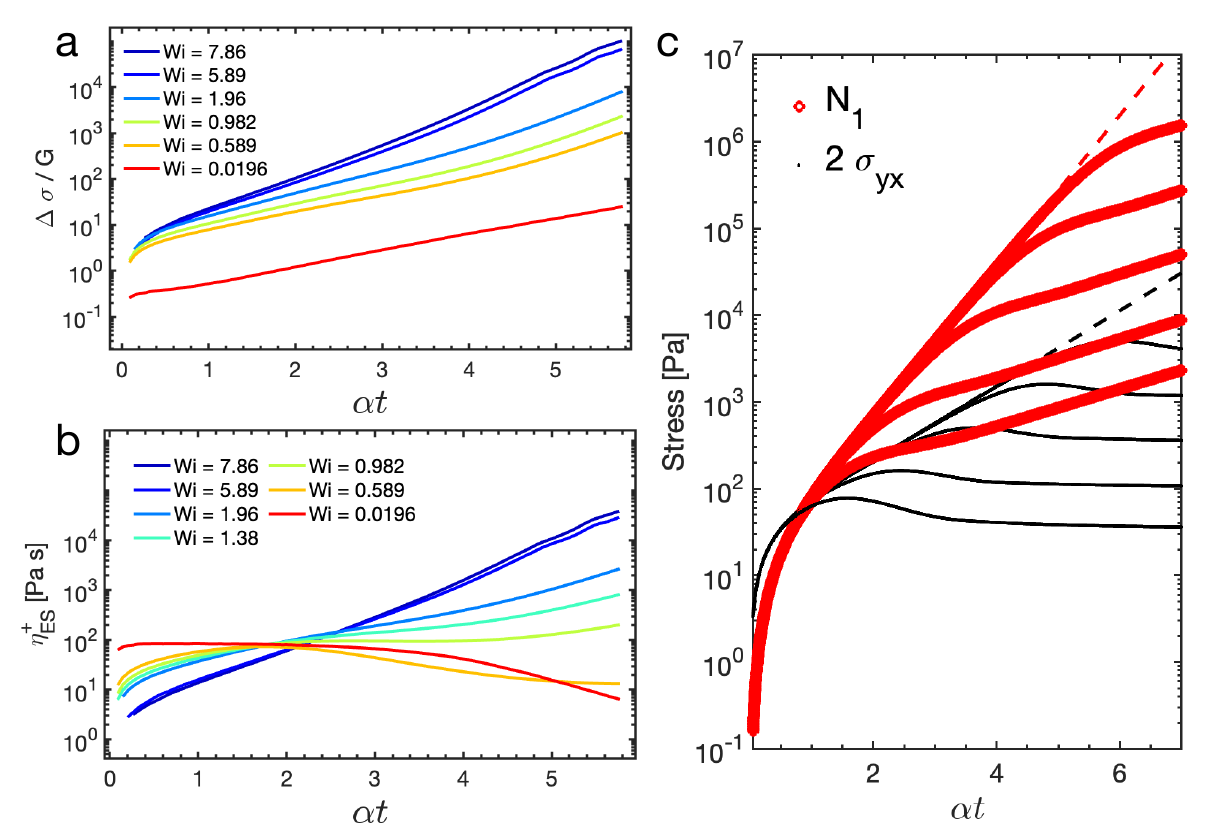}
    \caption{\label{fig:PIBsoln} Example: Shear Thinning PIB Fluid, (a) Normalized total principal stress (c)Transient extensional viscosity using$ \eta_s = 0.71$ Pa s and $\tau = 1.9645$s from experimental data in table \ref{tab:table1} c) Geisekus model in exponential shear for investigating effects of shear thinning in exponential shear flows. Mobility parameters shown are $a = 10^{-4}, 10^{-3}, 10^{-2}, 10^{-1}, 0.5$. $\alpha = 5, \tau = 1$. Further details in Appendix \ref{App:Geisekus}.}
\end{figure}

It is interesting (and slightly puzzling) that there does not appear to be substantial effect on the transient extensional viscosity due to shear thinning (especially at the larger $\alpha$ rates). To examine this, consider the synthetic exponential shear data (shown for Wi$ = 4$) for the Geisekus constitutive model(see Appendix B, \cite{larson2013constitutive}) which captures shear-thinning through a mobility parameter, $a$. When this mobility parameter is set to 0, and the second normal stress difference is set to 0, the dotted lines reproduce the Upper-Convected Maxwell solution. When the mobility parameter  starts to increase (from $10^{-4}$ to $0.5$) both the normal stress difference (in red) and the shear stress (in black) are affected. However, the shear stress is significantly more affected than the normal stress difference. Note that even for an aggressive mobility parameter of 0.5, the Normal stress difference still experiences stress growth at Hencky strain amplitudes of 2 to 6. At high Hencky strains, almost all of the information about the transient extensional viscosity is in the Normal stress difference. Figure \ref{fig:PIBsoln}c gives insight into this idea that the normal stress difference is far less affected by shear thinning than the shear stress. 

In principle, if one was intent on using this method on even more aggressively shear-thinning fluids, one \textit{could} construct a correction to the normal and shear stresses in the total principal stress (similar to the correction terms for solvent and inertial effects in section \ref{sec:corrections}) to account for the behavior. However, given such an estimated correction would require fitting to a model (unlike the other corrections), it is outside the scope of this current work -- and possibly unnecessary, given the findings in figures\ref{fig:PIBsoln}a-c. 

It is also worthwhile to observe that the crossing point where $N_1$ and $2 \sigma_{yx}$ occurs is early enough to be robust against most of the effects of shear thinning shown in figure \ref{fig:PIBsoln}c.

\subsection{\label{sec:level3} Demonstrating Finite Extensibility}
\begin{figure}
    \centering
    \includegraphics[width= 1.1 \linewidth]{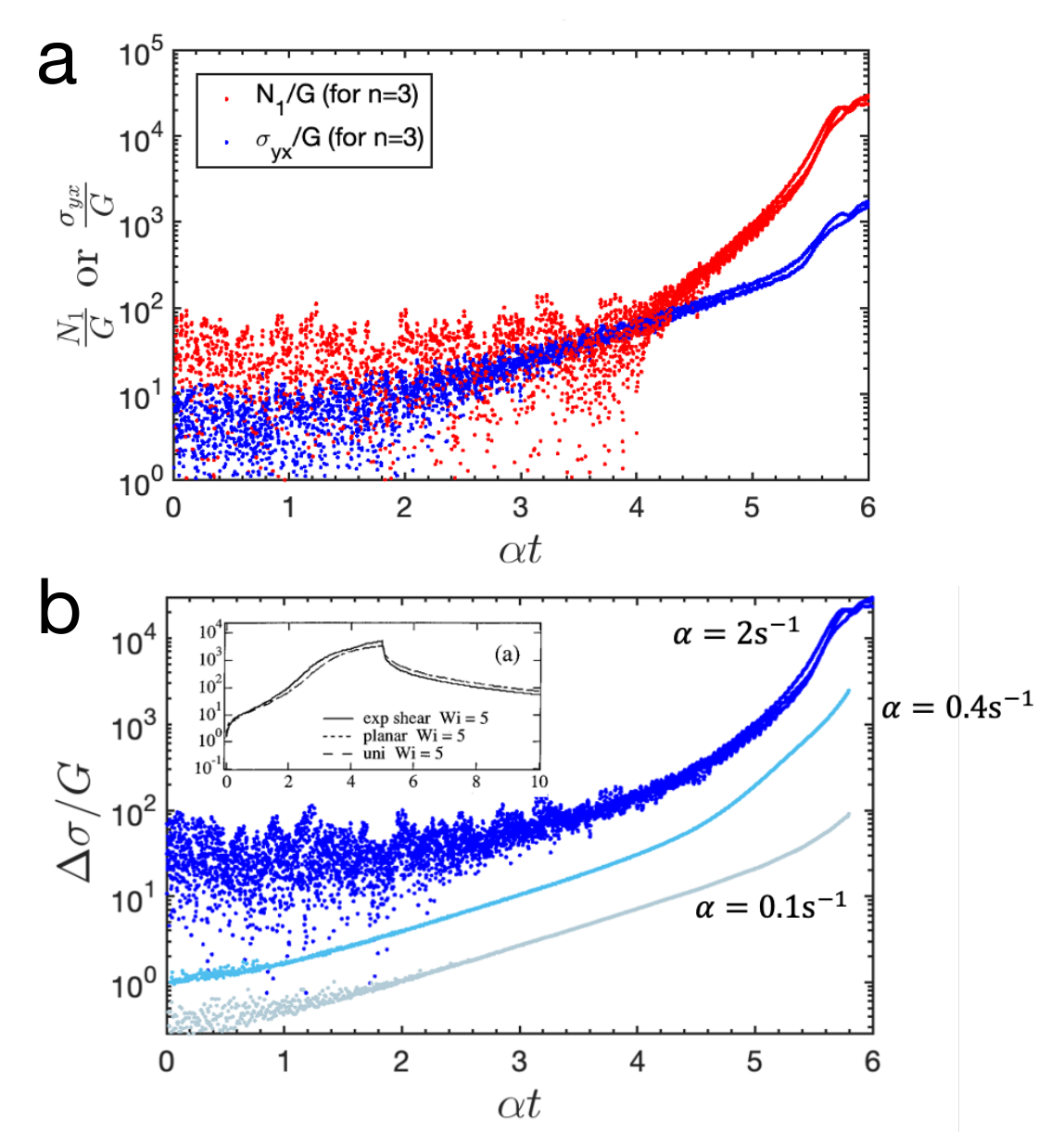}
    \caption{ \label{fig:finiteExtensibility}
By suppressing edge fracture with an immiscible, highly viscous oil at the rim of the sample, exponential shear can reach sufficiently high strain amplitudes in some fluids to approach finite extensibility. (a) Components of the stress (b) total principal stress (normalized by the modulus) for several different effective Hencky strain rates.  Note that for sufficiently high rates, the principal stress levels out, simular to the prediction (inset) from \citet{kwan2001experimental}. Here we used a 8mm parallel plate geometry with a very narrow gap ($h= 0.7$ mm), using an immiscible oil around the edge to suppress edge fracture. Video imaging was performed to ensure no sample ejection occurred prior to the end of the experiment.}
    \label{fig:enter-label}
\end{figure}
The PIB polymer solution studied in section \ref{sec:PIBFluid} exhibits finite extensibility at \textit{approximately} just beyond the maximum Hencky strain of the Ares G2 ($\varepsilon_{lim} \sim 5.8$ with a 25mm, 0.1 rad cone and plate). 

The normal stress transducer and shear rate max limit were at a maximum with that geometry (our lab's smallest cone and plate geometry). To extend the operational range of the instrument just slightly, we moved to an 8mm parallel plate with a very narrow gap (h = 0.7mm). Additionally, a viscous oil was used at the rim to delay edge fracture. Due to the non-viscometric flow caused by the parallel plate geometry, high-speed videos of each experiment were also simultaneously collected to ensure that edge fracture and sample ejection did not occur during the period of stress growth. It was common for the sample to be ejected \textit{after} the abrupt cessation at these extreme Hnecky strains, and important to reload a new sample after each run. 

Raw component data from three trials is shown in Figure \ref{fig:finiteExtensibility}a (with no filtering/smoothing, to be careful about minor effects to the shape of the data between $\varepsilon= 5$ and $6$). The normalized total principal stress is shown in figure \ref{fig:finiteExtensibility}b. A plateau near $\alpha t = \varepsilon = 5.85$ is observed, consistent with the extensional literature on approximately the Hencky strain we might expect for the onset of finite extensibility. It is noted that this data was collected at $\alpha = 2$ s$^{-1}$ (Wi $\sim 1.4$); a steady-extension condition that could only be tested on a Filament stretching rheometer or microfluidic device (not an elasto-capillary driven extensional rheometer).

\section{\label{sec:level1} Formal Experimental Validation: Comparing Exponential Shear with Extensional Rheometry}
\begin{figure*}[ht!]
\includegraphics[width=0.8\linewidth]{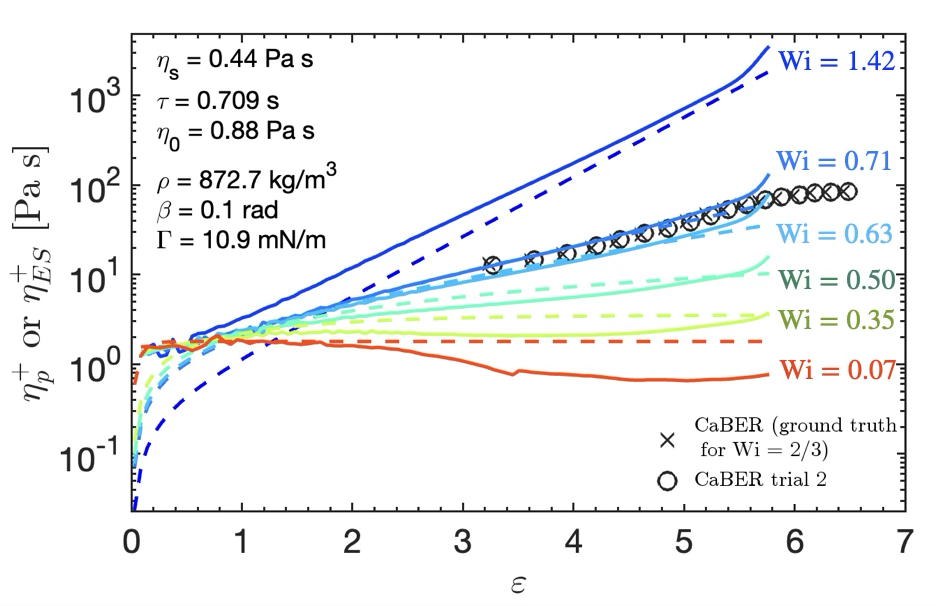}
\caption{\label{Punchline} Comparison of transient planar extensional viscosity for (1) exponential shear experiments at Wi $= 0.07$ to $1.42$ on a $0.3\%$ wt PIB Boger fluid, shown in solid lines (2) Planar extensional simulated viscosity using the upper convected maxwell model and (3) Capillary breakup extensional rheometery in black symbols.}
\end{figure*}

Next, we investigated if exponential shear can be quantitatively validated against extensional experiments, something that to our knowledge, never has been done before in exponential shear at intermediate Weissenberg number (likely due to lack of an appropriate material function). Even with this our new updated material function however, planar extensional validation proves challenging because of a lack of availability of commercially-available experimental methods for true, \textit{planar} extension that can reach Hencky strains of $\sim 6$. 

However, \textcite{kwan2001experimental} found that (simulated) planar vs uniaxial transient extensional viscosity are remarkably \textit{quantitatively} similar in exponential shear (see inset panel to Figure \ref{fig:finiteExtensibility}b). 

Leveraging this observation (transient extensional viscosity in planar extension versus uniaxial flows are practically equivalent), we performed uniaxial elastocapillary tests (CaBER) on the 0.3\% wt Boger fluid from section \ref{sec:PIBFluid}. The apparent extensional viscosity is given by \textcite{anna2001elasto}:
\begin{equation}
    \eta_{e,app} = -\frac{\Gamma}{\frac{d D(t)}{dt}}
\end{equation}

 The Hencky strain ($\alpha t$ in exponential shear) is equivalent to:
\begin{equation}
    \varepsilon = 2 \ln(R_0/R)
\end{equation}. 

Surface tension measurements were obtained on the DataPhysics DCAT25 tensiometer (using a Wilhelmy plate), and shown to be 10.9 mN/m. The density of the fluid was calculated to be $872.7$ kg/m$^3$. A high-speed camera (VisionResearch Phantom Miro) was used to collect the frames of the thinning neck diameter. The resulting effective extensional viscosity is shown in black symbols on figure \ref{Punchline}.    

The PIB Boger fluid data was processed with the new material function, using no fitting parameters or models -- only quantities directly from the experimental data in table 1 ($\eta_s = 0.44$ Pa s and $\tau = 0.709$ s). 

Additionally, we can compare these experiments to the simulated planar extension of a model UCM fluid (given by equation \ref{eq:planar}). These predictions are shown in dotted lines for each Wi. 

As shown in figure \ref{Punchline}, the extensional data falls precisely at Wi = 2/3, quantitatively validating the material function \textit{at intermediate Wi} against both true experimental extensional data (uniaxial) -- and planar extensional simulated data. 

\section{\label{sec:level1} MIT-StrongFlow: an open-source analysis tool}

In the spirit of MIT-LAOS (\textcite{ewoldt2009nonlinear}) and MIT-OWCh (\textcite{geri2018time}), we will attach an open software analysis tool is currently under development to help facilitate the processing of data using this material function (including implementing the necessary corrections in section \ref{sec:corrections} and automating the parameter extraction for $\tau$ and $\eta_s$). These codes will be openly available on github, once the manuscript concludes the review process. Please contact the authors if you need these codes urgently, and we can provide sooner upon reasonable request.

\begin{acknowledgments}
The authors would like to thank Dr. Eugene Pashkovski of Lubrizol Corporation for providing many of the PIB materials used in this study. We would also like to thank Dr. Rishabh More for discussions on the prior characterization of theses materials and illuminating discussions about the onset of elastic instability.  We wish to acknowledge the support of Motif Foodworks Inc. in providing financial support in 2022-2024 for this study.
\end{acknowledgments}

\section*{Author Contributions Statement}
GHM and LAK performed the original research in this paper. LAK wrote the original manuscript draft. GHM and LAK originated the ideas in this paper. 

MWB, RAN and SKB provided industrial postdoctoral funding to MIT for LAK for a portion of this project. MB, SK, and GHM originally conceptualized an implementation of a cyclical version of exponential shear on food materials; the topic of a separate paper with no conceptual overlap (but which ended up \textit{requiring} this study and material function to proceed with). 

\section*{Data Availability Statement}

The data that support the findings of this study are available from the corresponding author upon reasonable request.

\appendix

\section{Linear Rheological Characterization of test fluids: Using 5-mode Oldroyd-B fit (viscosity weighted) and Rouse-Zimm}

To validate the statement $\btau \approx \tau_{exp} = f(t^\star)$, we need a good understanding of the true value of $\btau$ using standard techniques -- preferably using a more sophisticated model than Upper Convected Maxwell. To achieve this, we performed linear viscoelastic measurements on each of the polymeric fluids in table 1, and fit a 4-mode Oldroyd-B model to these fluids using the data (i.e. $G^*(\omega) = i\,\omega\,\eta_s + \sum_{i=1}^{4} \frac{G_i\, (i\,\omega\,\tau_i)}{1 + i\,\omega\,\tau_i}
$). At risk of being verbose, we can write this explicitly as: 
\begin{equation}
    G'(\omega) = \sum_{i=1}^{4} \frac{G_i\, (\omega\,\tau_i)^2}{1 + (\omega\,\tau_i)^2}
\end{equation}
and 
\begin{equation}
    G''(\omega) = \omega\,\eta_s + \sum_{i=1}^{4} \frac{G_i\, (\omega\,\tau_i)}{1 + (\omega\,\tau_i)^2}
\end{equation}
We used the sum of squared residuals as the cost function. Root mean squared error (RMSE) is computed and reported in the caption for each fluid with the parameters.

\begin{table}[ht!]
    \centering  
\begin{tabular}{lccc}
\hline
Mode & $G_i$ (Pa) & $\tau_i$ (s) & $\eta_{p,i} \mathrm{(Pa}\cdot\mathrm{s)}$ \\
\hline
1 & 0.582 & 0.215   & 0.125 \\
2 & 8.22  & 0.00635 & 0.0522 \\
3 & 1.88  & 0.0321  & 0.0601 \\
4 & 0.0165& 3.43    & 0.0566 \\
5 & 1.14e-06 & 4.91 & 5.59e-06 \\
\hline
\end{tabular}
\label{tab:fitted_params}
    \label{tab:my_label}
    \caption{$0.3$\% wt PIB Boger Fluid: 5-mode Old-B fit}
\end{table}


We report the viscosity-weighted average relaxation time in table 1 as a comparison for $\tau_{exp}$.  
\begin{equation} 
\bar{\tau} = \frac{\sum_{i=1}^N G_i \,\tau_i^2}{\sum_{i=1}^N G_i \,\tau_i}
\end{equation}

For a better understanding of how close this timescale is to a different model (with very different / more molecular underpinnings) we also extract the Rouse-Zimm average relaxation time, $\tau_z$ to compare. This Fit is also shown (for the 0.3 percent wt BF) in Figure(appendix fig). 

\begin{figure}
    \centering
    \includegraphics[width=\linewidth]{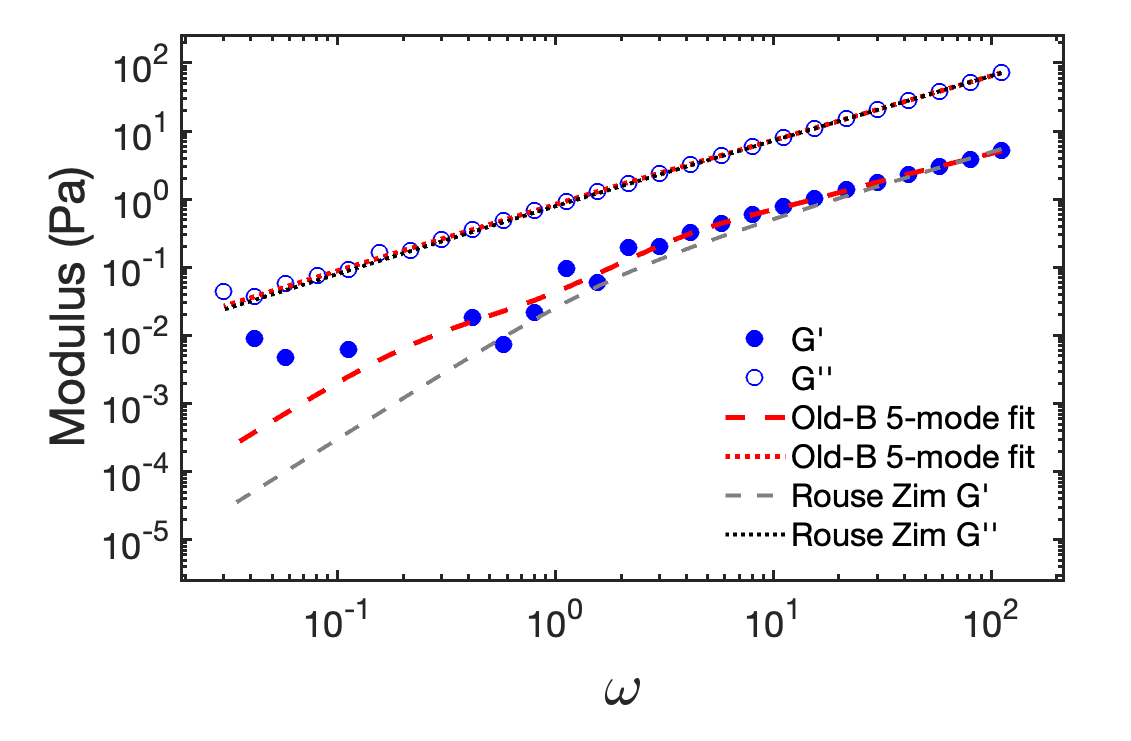}
     \caption{Experimental Data vs model fits for 5 modes of Oldroyd-B for the 0.3 \% wt PIB Boger Fluid. Viscosity-weighted relaxation time for Oldroyd-B is $0.76$ s. RMSE between Oldroyd-B model and data was computed to be 0.0294 Pa. For Rouse-Zimm, the solvent quality $\nu = 0.3504$ and prefactor $\frac{\phi \, K \, t}{b^3 \, n} = 1.03$. RMSE error reported for Rouse Zimm is $0.111$ Pa. }
    \label{fig:enter-label}
\end{figure}

\begin{table}[ht!]
    \centering  
    \begin{tabular}{lccc}
        \hline
        Mode & $G_i$ (Pa) & $\tau_i$ (s) & $\eta_{p,i}$ (Pa$\cdot$s) \\
        \hline
        1 & $3.74\times10^{4}$ & $0.000297$ & $11.1$ \\
        2 & $2.12\times10^{3}$ & $0.00465$  & $9.86$ \\
        3 & $17.1$            & $0.0409$   & $0.701$ \\
        4 & $1.51$            & $0.134$    & $0.202$ \\
        5 & $1.04$            & $0.134$    & $0.139$ \\
        6 & $2.92$            & $0.134$    & $0.391$ \\
        \hline
    \end{tabular}
    \caption{PDMS: 6-mode Old-B fit.}
    \label{tab:my_label}
\end{table}

\begin{figure}
    \centering
    \includegraphics[width=\linewidth]{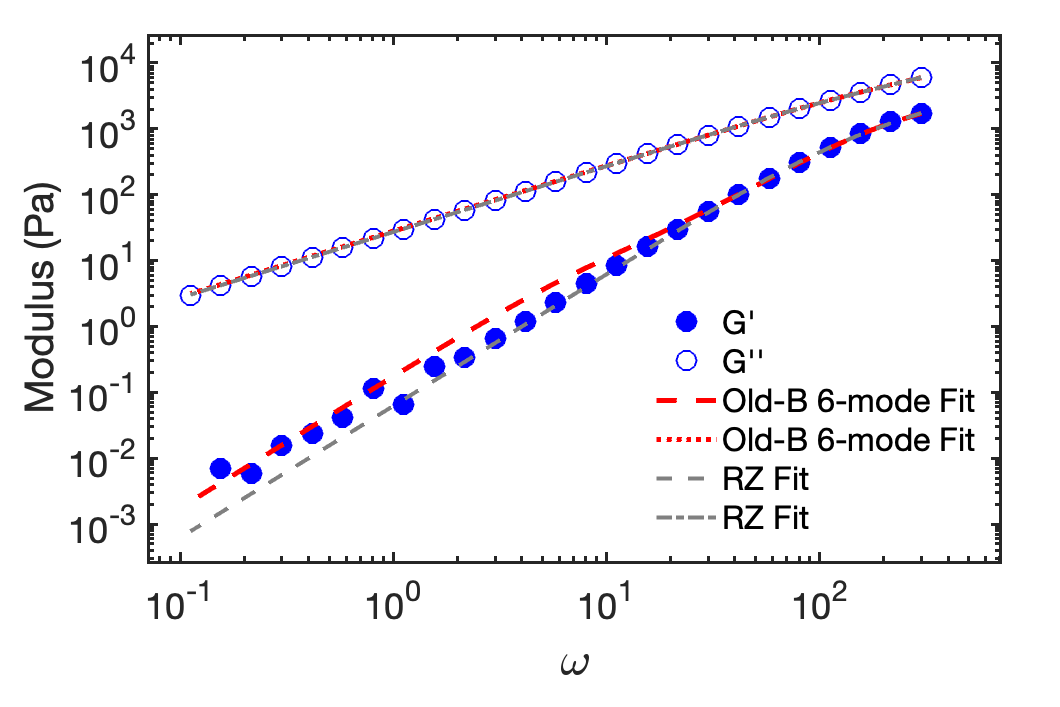}
    \caption{Experimental Data vs model fits for 6 modes of Oldroyd-B for the PDMS fluid. Fit parameters in table IV. }
    \label{fig:enter-label}
\end{figure}

Orange BF:
\begin{table}[ht!]
    \centering  
    \begin{tabular}{lccc}
        \hline
        Mode & $G_i$ (Pa) & $\tau_i$ (s) & $\eta_{p,i}$ (Pa$\cdot$s) \\
        \hline
        1 & 20.2  & 0.0095 & 0.192 \\
        2 & 5.57  & 0.0586 & 0.326 \\
        3 & 2.32  & 0.257  & 0.598 \\
        4 & 1.06  & 0.758  & 0.800 \\
        5 & 0.792 & 2.80   & 2.22  \\
        \hline
    \end{tabular}
    \caption{Orange shear-thinning fluid: 5-mode Oldroyd-B fit. Fitted solvent viscosity: 0.238~Pa$\cdot$s, viscosity-weighted average relaxation time: 1.69~s, RMSE between model and data: 0.26~Pa.}
    \label{tab:newfluid}
\end{table}

\begin{figure}
    \centering
    \includegraphics[width=\linewidth]{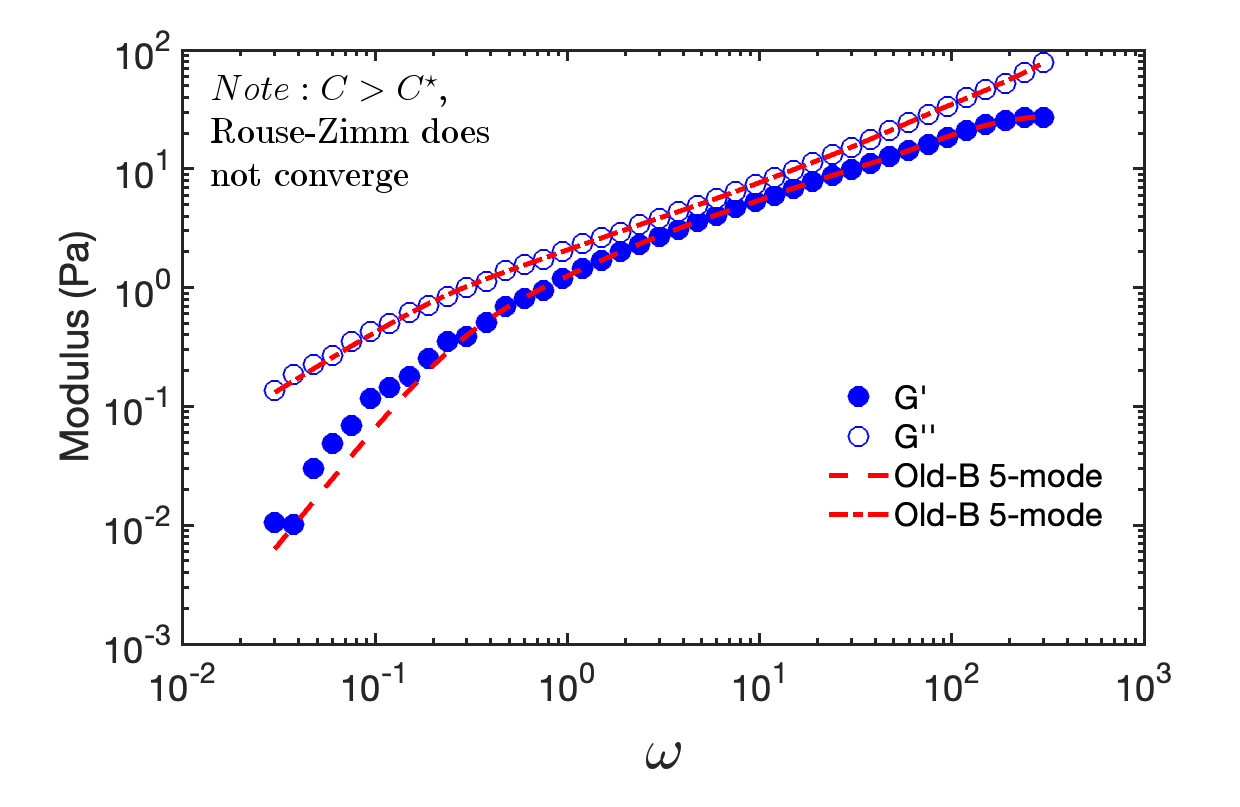}
    \caption{Experimental Data vs model fits for 5 modes of Oldroyd-B for the 3\% wt. shear thinning PIB fluid. Fit parameters in table IV. Because the polymer concentration is so high, Rouse-Zimm is not an appropriate alternative model. Fitting code did not converge.}
    \label{fig:enter-label}
\end{figure}

\begin{figure}
    \centering
    \includegraphics[width = \linewidth]{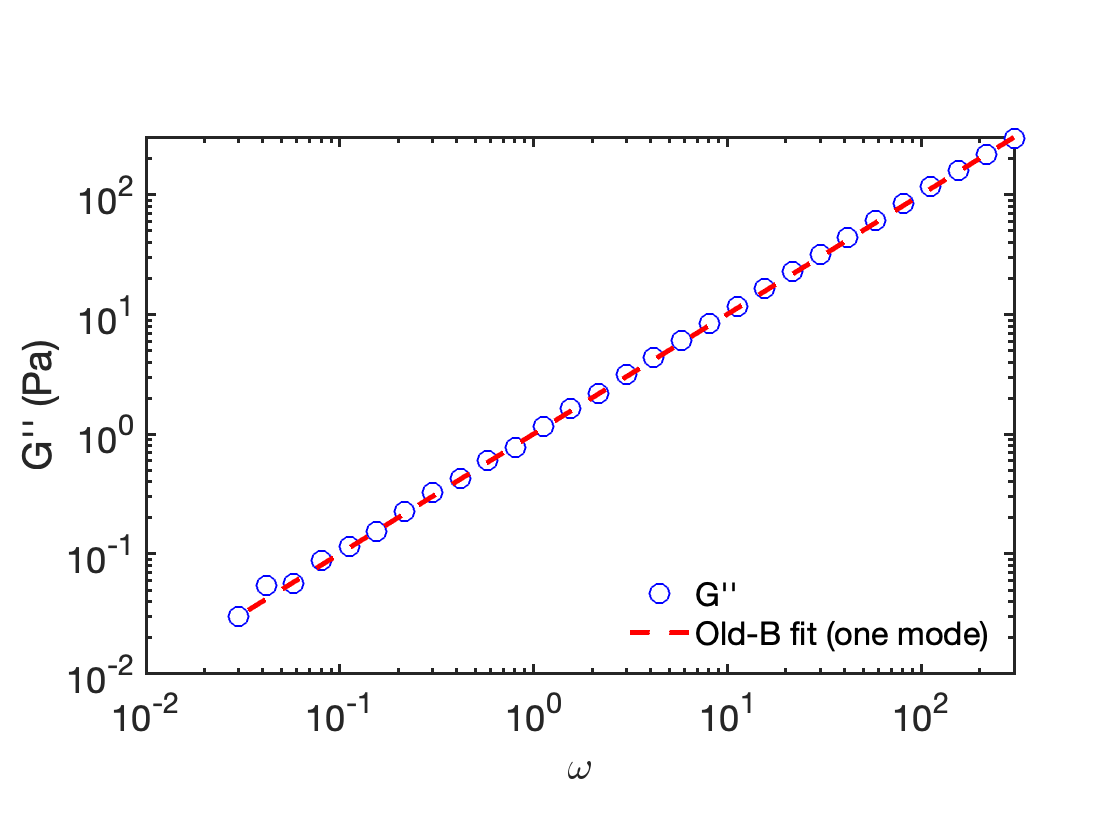}
    \caption{Viscosity-standard 1 Pa-s silicone oil}
    \label{fig:enter-label}
\end{figure}

\section{Geisekus model}
\label{App:Geisekus}
The constitutive equation for the Geisekus model is given by:

\begin{eqnarray*}  
\label{eq:1}
\boldsymbol{\sigma}
\;+\;
\lambda_{1}\,\overset{\nabla}{\boldsymbol{\sigma}}
\;+\;
a\,\frac{\lambda_{1}}{\eta_{0}}
\bigl\{\boldsymbol{\sigma} \cdot \boldsymbol{\sigma}\bigr\}
\;-\;
a\,\lambda_{2}
\bigl\{\dot{\boldsymbol{\gamma}} \cdot \boldsymbol{\sigma}
       \;+\;
       \boldsymbol{\sigma} \cdot \dot{\boldsymbol{\gamma}} \bigl\}
\;= \\\;
\eta_{0}\,\Bigl[\dot{\boldsymbol{\gamma}}
       \;+\;
       \lambda_{2}\,{\ddot{\boldsymbol{\gamma}}}
\;-\;
a\,\frac{\lambda_{1}^{2}}{\lambda_2}
\Bigl[\dot{\boldsymbol{\gamma}} \cdot \dot{\boldsymbol{\gamma}}
      \;\Bigr]    
\end{eqnarray*}

These equations describe the shear-thinning behavior of the fluid under the Geisekus model. This model was used to construct the synthetic data shown in figure \ref{fig:PIBsoln}c, demonstrating that the first normal stress difference is \textit{far less} affected by a positive shear thinning exponent than the shear stress (which effectively completely suppressed at large Hencky strain amplitudes). This model was used to numerically simulate the stresses under a strong exponential shear (Wi = 5). 



\section{Persistence of Strain}
\begin{figure*}
\label{PofStrain}
    \centering
    \includegraphics[width=\linewidth]{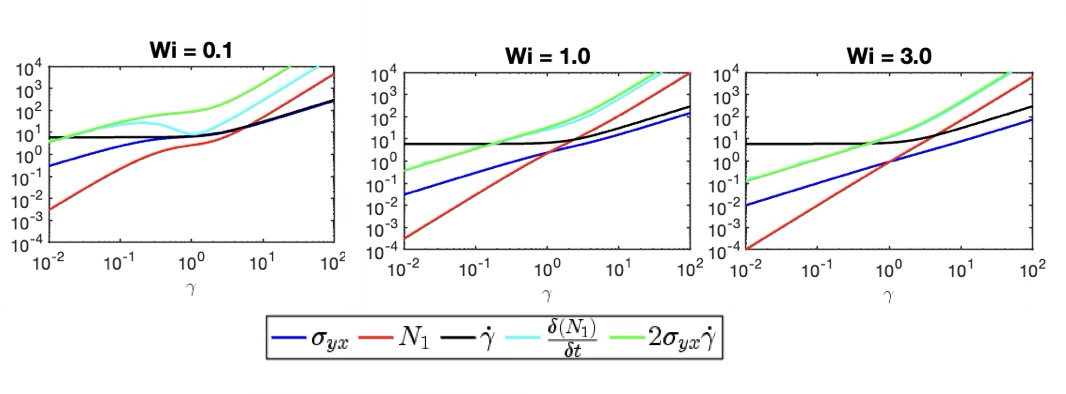}
    \caption{A brief note on persistence of strain: Here we show synthetic Upper convected Maxwell data showing the relative growth of the first normal stress difference compared with the shear rate. Note that the Normal stress grows faster than the vorticity.}
    \label{fig:enter-label}
\end{figure*}

As a brief side note on persistence of strain: we show in figure \ref{PofStrain} that the first normal stress difference grows \textit{faster} than the strain rate (vorticity scales with strain rate to the first power) at any Weissenberg number -- evidence that if the total principal stress is considered (rather than the shear stress), this method can be used as an effective stretching flow method (as $N_1$ scales as strain rate to the second power).      


\section*{References}

\nocite{*}
\bibliography{aipsamp}

\end{document}